\documentclass[12pt]{iopart}
\usepackage{geometry}
\usepackage{graphicx}
\expandafter\let\csname equation*\endcsname\relax
\expandafter\let\csname endequation*\endcsname\relax
\usepackage{amsmath}
\usepackage{breqn}
\usepackage{autobreak}
\usepackage{color}
\emergencystretch=\maxdimen
\allowdisplaybreaks

\usepackage[colorlinks=true,bookmarks=false,citecolor=blue,urlcolor=blue]{hyperref}
\usepackage{setspace,cite}

\newcommand{\bea}{\begin{eqnarray}}
\newcommand{\eea}{\end{eqnarray}}
\newcommand{\bes}{\begin{subequations}}
\newcommand{\ees}{\end{subequations}}
\newcommand{\ds}{\displaystyle}

\begin{document}

\title[Painlev\'e Analysis and Higher-Order Rogue Waves $\dots$]{Painlev\'e Analysis and Higher-Order Rogue Waves of a Generalized (3+1)-dimensional Shallow Water Wave Equation}

\author{Sudhir Singh$^{1}$, K. Sakkaravarthi$^{2,3,4}$, T. Tamizhmani$^{5}$ \&  K. Murugesan$^{1}$}

\address{$^{1}$Department of Mathematics, National Institute of Technology, Tiruchirappalli -- 620 015, Tamil Nadu, India\\ $^{2}$Young Scientist Training Program, Asia-Pacific Center for Theoretical Physics, POSTECH Campus, Pohang -- 37673, Republic of Korea\\$^{3}$PG \& Research Department of Physics, Bishop Heber College (Autonomous), Affiliated to Bharathidasan University, Tiruchirappalli -- 620 017, Tamil Nadu, India\\ $^{4}$Centre for Nonlinear Dynamics, School of Physics, Bharathidasan University, Tiruchirappalli -- 620 024, India\\ $^{5}$Department of Mathematics, School of Advanced Sciences, Vellore Institute of Technology, Vellore -- 632 014, India}

\ead{ksakkaravarthi@gmail.com (Corresponding author)}

\setstretch{1.25}

\begin{abstract}
Considering the importance of ever-increasing interest in exploring localized waves, we investigate a  generalized (3+1)-dimensional Hirota-Satsuma-Ito equation describing the unidirectional propagation of shallow-water waves and perform Painlev\'e analysis to understand its integrability nature. We construct the explicit form of higher-order rogue wave solutions by adopting Hirota's bilinearization and generalized polynomial functions. Further, we explore their dynamics in detail, depicting different pattern formation that reveal potential advantages with available arbitrary constants in their manipulation mechanism. Particularly, we demonstrate the existence of singly-localized line-rogue waves and doubly-localized  rogue waves with multiple (single, triple, and sextuple) structures generating triangular and pentagon type geometrical patterns with controllable orientations that can be altered appropriately by tuning the parameters. The presented analysis will be an essential inclusion in the context of rogue waves in higher-dimensional systems.
\end{abstract}

%

\setstretch{1.20}
\section{Introduction}
	Rogue waves are interesting localized nonlinear wave structures that ``appear from nowhere and disappear without any trace"  \cite{aa8} result into a huge impact in the systems they emerge for the past two decades. They become very important among several nonlinear entities and attract much focus of the researchers working across different branches of science such as optics, Bose-Einstein condensate, plasma, and oceanography \cite{aa7,aa9,aa10,aa11,aa12,aa13,aa14}. These rogue waves are high amplitude unstable structures, short-lived, and localized in both space and time, which manifest them to be a stand-alone among different localized nonlinear waves. 
	Note that, apart from the rogue waves, there exist several other types of nonlinear coherent structures that are prevalent because of their mathematical beauty and tremendous applications. To name a few, solitons, solitary waves, lumps, breathers, peakons, compactons, dromions, solitoffs, ring and loop solitons, foldons (folded solitary waves), periodic waves and their interactions are of potential interest \cite{yang-book,aa2,aa1,aa3,aa4}. 
	These waves have enriched beauty and do occur in various integrable and non-integrable nonlinear dynamical systems. A few of these localized waves are stable over a long distance and well established, for example solitons that are exponentially localized wave solutions and preserve their identities even after collision with other solitons. Due to such remarkable stability and intriguing collision properties solitons perceive a prominent role in almost all areas and they are being studied rigorously over the past fifty years with proven facts of multifaceted applications \cite{soliton,soliton2}. 
	In contrast to the solitons, various other waves are highly unstable but possess vibrant dynamical features. For example, lumps are rational analytic function solutions localized in all spatial directions and they can be reduced by the long-wave limit of $N$-soliton solutions \cite{wt}. Their interaction with solitons is turn out to be both elastic and inelastic in certain soliton equations \cite{aa5}. It is important to note that rogue waves are described by rational solutions, but localized both in space as well as in time, pertinent to certain indeterministic behaviour and carry different names (such as as abnormal waves, freak waves, monster waves, killer waves, giant waves and extremes waves) due to their threatening nature 
	\cite{aa8,aa7,aa9,aa10,aa11,aa12,aa13,aa14}. Except for the solitons and rogue waves, other mentioned localized wave structures of nonlinear partial differential equations are usually extracted either as special cases of soliton solutions or using particular test functions and general approaches are not much developed for such nonlinear structures.
	
	The very first report on rogue wave was in the year 1983 described as a rational analytical solution to the celebrated nonlinear Schr\"odinger equation by H. Peregrine and referred to as Peregrine breather/soliton after his name \cite{aa16}. 
	For a detailed review on the experimental and theoretical investigations can be found in a special collection \cite{aa15,aa17,jpa2017}. Apart from these, it is necessary to highlight some important studies on rogue waves in one- and higher-dimensional nonlinear models. Especially, several (1+1)-dimensional water wave models starting from the renowned Korteweg-de Vries (KdV) equation, Boussinesq equation, Ito equation, nonlinear Schr\"odinger equation and Benjamin-Ono equation, along with their (2+1) and (3+1)-dimensional integrable/non-integrable family of models, including the celebrated Kadomtsev-Petviashvili (KP) equations, which is nothing but the KdV equation in (2+1)-dimensions, are to name a few \cite{aa2,aa1,aa3,aa4,wt,aa5}. 
	The higher dimensional analogue of these equations are well suited for their well-grounded behaviour and it allows the emergence of physically important localized structures. Though there exists tremendous amount of works on rogue waves based on different theoretical and experimental investigations pertaining to their generation mechanism and dynamical behavior of rogue waves in the past decade, still they remain to be a debatable subject with much enthusiasm \cite{aa15,aa17,jpa2017}. Though the dynamics is continue to be fascinating, the higher dimensional nonlinear models are tough to explore mathematically. 
	Among the many difficulties, analysing the integrability nature becomes one of the important questions next to obtaining their solutions. The classical KdV equation is integrable in the sense of Painlev\'e analysis, possesses lax pair and infinite conserved quantities, solvable through Inverse scattering transform and bilinear formalism, and exhibits multi-soliton solution. However, many of its generalized models fail to have all/any of these exciting properties. These properties can be possible partially, for example, Painlev\'e integrable models may not have a lax pair and vice-versa the models admitting lax pair may not be Painlev\'e integrable. So, studies on the integrability and localized structures of higher-dimensional equations are important aspects of study for a complete understanding of the associated nonlinear systems \cite{ML-book}. 
	
	Motivated by the importance of rogue waves and higher dimensional models, our aim of the present work is to investigate the integrability nature and dynamics of higher-order rogue waves in the following (3+1)-dimensional generalized Hirota-Satsuma-Ito equation:
	\begin{equation}
	\Gamma_1 [3(u_x u_t)_x+u_{xxxt}]+\Gamma_2[3(u_x u_y)_x+u_{xxxy}]+\Gamma_3 u_{yt}+ \Gamma_4 u_{xx}+ \Gamma_5 u_{xy}+ \Gamma_6 u_{xt}+ \Gamma_7 u_{yy}+\Gamma_8 u_{zz}=0, \label{eq13}
	\end{equation} 
	where $x,y,z$ represent three spatial dimensions while $t$ denotes time and $\Gamma_j, j=1,2,3,\dots,8,$ are arbitrary constants defining the magnitude of dispersive and nonlinear characteristics that can reflect significant physical utility in the associated nonlinear waves. Physically, the above considered HSI model (\ref{eq13}) describes an unidirectional propagation and interactions of surface waves in shallow water \cite{ck}. The difference between the coefficients $\Gamma_1$ and $ \Gamma_2$ is that the former is characterized by both spatial and temporal effects while the latter is characterized by only spatial coordinates with combined linear and nonlinear dispersion-nonlinearity contributions. These two terms affect and control the dispersion relation pertaining to Eq. (\ref{eq13}), and so the wave number will further complicate the  evolution dynamics of the considered model \cite{ck}. 
	
	The parameters $\Gamma_j$ appearing in the above generalized (3+1)D HSI model play an important role and for different choices of these $\Gamma_j$ parameters Eq. (\ref{eq13}) reduces to several nonlinear wave equations reported recently which contain interesting results on different nonlinear wave solutions and we list a few for completeness. 
	(i) Resonant multi-soliton solutions of Eq. (\ref{eq13}) with $\Gamma_1=1$, $\Gamma_2=0$, $\Gamma_3=\delta_1$, $\Gamma_4=\delta_2$, $\Gamma_5=\delta_3$, $\Gamma_6=\delta_4$, $\Gamma_7=\delta_5$, $\Gamma_8=0$ are obtained using linear superposition principle and bilinear form \cite{ck}. For the same model, breather wave solutions are discussed using bilinear B\"acklund transformation \cite{m1}. 
	(ii) Interaction waves and lump solutions are constructed using the Hirota bilinear form \cite{wx} and lump solution using Bell polynomial \cite{epjp20} for Eq. (\ref{eq13}) with $\Gamma_1=\Gamma_3=\Gamma_4=1$, $\Gamma_i=0$, $i=2,5,6,7,8$. 
	(iii) $N$-soliton and hybrid wave solutions at a long wave limit are reported for Eq. (\ref{eq13}) when $\Gamma_1=c_1, \Gamma_3=c_2, \Gamma_4=c_3, \Gamma_6=c_4$, $\Gamma_i=0$, $i=2,5,7,8$ \cite{zz}. 
	On the other hand, equation (\ref{eq13}) reduces to (iv) generalized
	Calogero-Bogoyavlenskii-Schiff equation \cite{st, lh} for $\Gamma_2=1$, $\Gamma_5=\delta_1, \Gamma_6=1, \Gamma_7=\delta_2$, $\Gamma_i=0, i=1,3,4,8$, where localized lump solutions using Hirota Bilinear formalism with quadratic polynomial test function are obtained \cite{st}, while breather and interaction solutions are obtained using homoclinic (two wave and four wave) test functions \cite{lh}, (v) dimensionally reduced general Jimbo-Miwa equation \cite{hw} for $\Gamma_2=1$, $\Gamma_3=-2,  \Gamma_4=-3, \Gamma_6=4$, $\Gamma_i=0, i=1,5,7,8$ with lump solutions and their interactions with one stripe soliton and rogue waves, (vi) (3+1)D generalized KP equation for $\Gamma_2=1$, $\Gamma_3=1,  \Gamma_6=1, \Gamma_8=-1$, $\Gamma_i=0, i=1,4,5,7$ \cite{wx2} with Wronskian and Grammian solutions using Hirota bilinear method with the help of P\"lucker relation and the Jacobi identity for determinants, (vii) (3+1)D generalized BKP equation for $\Gamma_2=1$, $\Gamma_3=-1, \Gamma_4=-6, \Gamma_8=3$, $\Gamma_i=0, i=1, 5,6,7$  \cite{js}, with soliton solutions in Wronskian form using bilinear formalism, and another case (viii) $N$-solitary wave, homoclinic breather, and rogue wave solutions of (3+1)D nonlinear wave equation when $\Gamma_2=1$, $\Gamma_3=-1,  \Gamma_4=3, \Gamma_8=-3$, $\Gamma_i=0, i=1,5,6,7$ \cite{mj}. 
	Apart from the above listed models, different class of higher-dimensional nonlinear equations under various physical settings have been reported with interesting results on rogue waves in the recent years, see for example \cite{cnsns2020,wx1,nld1,nld2,nld3,nld4,nld5,he21,ps21,nld6}. Moreover, under the vanishing effect of the parameter $\Gamma_2=0$, the model (\ref{eq13}) comes under the family of Hirota-Satsuma equations.  {{ Recently, another (2+1)D Hirota-Satsuma-Ito equation is also investigated and several interesting nonlinear waves are reported including multiple lumps, lump-solitary waves and lump-periodic waves in Ref. \cite{cjp2026}, which can be reduced from the present (3+1)D HSI model (\ref{eq13}) for $\Gamma_2=0, \Gamma_1=1$ and considering spatial-temporal transformation $z+t=T$.}} Moreover for the vanishing  $\Gamma_1$ parameter (\ref{eq13}) reduces to different soliton equations including Calogero-Bogoyavlenskii-Schiff, KP, BKP and  Jimbo-Miwa equations as mentioned above. From these reports, one can understand that the considered (3+1)D HSI equation (\ref{eq13}) is more general with much physical importance. 
	
 {{	Along with these lines,  the Hirota–Satsuma coupled KdV equations are also much-celebrated soliton models. Recently, the multisolitons and  the dynamics property for three-component Hirota–Satsuma coupled KdV equation is studied in  \cite{cjp2024}. The residual power series method is utilized to extract analytical solutions in \cite{cjp2027}, and a semi-analytical approach is used to generate the solutions of Hirota–Satsuma coupled KdV equations \cite{cjp2025}. Also, recently to understand the evolutionary dynamics of nonlinear waves completely, dual-mode Hirota–Satsuma coupled KdV equations is proposed and studied in \cite{cjp2022}. Recently, to dig deep into the whole dynamics of nonlinear waves, fractional counterpart soliton models have been proposed and studied. The fractional Hirota–Satsuma coupled KdV equations also gained much attention; several researchers studied the analytical solutions using different ansatz approaches \cite{cjp2023, cjp2028, cjp2029}. }}

	Based on the above perspectives, there arise several motivations to study the considered (3+1)D HSI equation (\ref{eq13}) for its integrability aspect with respect to infinite conserved quantities, Lax pair, inverse scattering transform, Hamiltonian formulations etc. and nonlinear wave solutions including solitons, solitary waves, periodic waves, breathers, lump structures, etc. However, in this work, we limit our objective to pursue only its integrability nature through Painlev\'e analysis and the dynamics of rogue waves. Particularly, we are interested to construct higher order rogue wave solutions based on Hirota bilinear formalism and generalized polynomial type seed solutions \cite{hi,pa,ss,zh,yang21} and to explore their dynamics through a detailed analysis on the effect of $\Gamma_j$ parameters. Without stretching the introduction much further, we shall proceed to implement Painlev\'e analysis and construct rogue wave solutions of the (3+1)D HSI model (\ref{eq13}). 
	
	
	The remaining part of this article is organised as follows.
	The Painlev\'e test is performed to study the integrability of the model (\ref{eq13}) in Sec. 2. Section 3 explains the methodology to extract the higher-order rogue wave solutions. The construction of explicit first-, second-, third-, and generalized $N$th-order rogue wave solutions along with a detailed discussion on their evolution dynamics in Sec. 4.  {{Section 5 briefly highlights the important results obtained in the work and future perspectives in a nutshell.}} The final section is allotted for conclusions derived from the present work.
	
	\section{Painlev\'e Integrability Analysis}
	Painlev\'e singularity structure analysis is one of the efficient tools to understand the integrability nature of any dynamical (ordinary/partial differential) equation in both one- and higher-dimensions \cite{pain-krus,aaa6,pain-kmt,sg1,pain-tk,pain-tk2,pain-tk3}. This includes three important steps such as the identification of leading order, determination of resonances and arbitrary analysis to ensure the availability of required number of free parameters. To perform the Painlev\'e test of the (3+1)D HSI equation (\ref{eq13}), first we need to identify the leading order by assuming the initial form as 
	\begin{equation}
	u \approx u_0 \phi ^{\alpha}, \label{peq1}
	\end{equation} 
	where $\alpha$ is negative integer to be determined, while $u_0$ and $\phi$ are analytic functions of $x,y,z$ and $t$. On substituting (\ref{peq1}) into equation (\ref{eq13}) and balancing the most dominant terms, we find that the leading order arise for $\alpha=-1$, for which the resultant leading order equation is obtained as 
	\begin{equation}
	(\Gamma_1 \phi _t+\Gamma _2 \phi _y ) ( u_0 -2 \phi _x) \phi _x^2 =0. \label{peq2} 
	\end{equation} 
	The next step is to find the resonance, which utilizes the full Laurent series with the known leading order $\alpha=-1$,
	\begin{equation}
	u=  \sum _{j=0} ^{\infty} u_j \phi ^{j+\alpha} \Rightarrow u_0 \phi ^{-1}+ \sum _{j=1} ^{\infty} u_j \phi ^{j-1}. \label{peq3}
	\end{equation}
	From the (3+1)D HSI equation (\ref{eq13}) with the help of above $u$ form (\ref{peq3}) and on equating the coefficient of $\phi ^{j-5}$ in the leading order, we get the following polynomial equation in $j$:
	\begin{equation}
	j^4 - 10 j^3 + 23 j^2 + 10 j - 2j=0. \label{peq4}
	\end{equation}
	Solution of the above equation (\ref{peq4}) leads to the required resonances and are found to be $j=-1,1,4,6$. As all of these  resonance values are integers, they indicate the possibility of equation (\ref{eq13}) to be integrable. But, for its confirmation one has to obtain required number of arbitrary function at each of these resonances. It is quite natural to note that the negative  resonance ($j=-1$) corresponds to the arbitrariness of the singular manifold $\phi (x,y,z,t)=0$.
	
	As mentioned before, the third and final step in the Painlev\'e test is the arbitrary analysis, where the condition for sufficient number of arbitrary functions at each values of resonance are evaluated and if so the equation can be confirmed as Painlev\'e integrable, otherwise the model will be identified as non-integrable in the Painlev\'e sense. For this purpose, by truncating the Laurent series (\ref{peq3}) up to the highest resonance value ($j=6$) as $u= u_0 \phi ^{-1}+ u_1+u_2\phi +u_3\phi^2 +u_4\phi^3 +u_5\phi^4 +u_6\phi^5$, we express the considered equation (\ref{eq13}) and look for arbitrariness of functions $u_j$ arising at different orders (coefficients) of $\phi$. 
	
	From the coefficient of $\phi^{-5}$, we obtain single equation for $u_0$, which is exactly same to that of leading order equation (\ref{peq2}) or simply $u_0=2 \phi _x$, and it shows that $u_0$ is not an arbitrary function. 
	Next, on collecting the coefficient of $\phi^{-4}$  corresponds to the resonance value $j=1$, we identify that  the resultant expression vanishes, which confirms the arbitrariness of $u_1(x,y,z,t)$ at $j=1$ as required. 
	To proceed further and for simplification, we adopt a Kruskal ansatz $\phi (x,y,z,t)= x+ \psi (y,z,t)$ as recommended in \cite{pain-krus,aaa6}. Analysing the coefficients of $\phi ^{-3}$ and $\phi ^{-2}$, respectively, we get the following explicit expressions for $u_2(x,y,z,t)$ and $u_3(x,y,z,t)$:
	\bes \bea
	&&u_2=-\frac{\psi_t(\Gamma_6+\Gamma_3 \psi_y+3\Gamma_1 u_{1,x}) +\psi_y(\Gamma_5+\Gamma_7 \psi_y+3 \Gamma_2 u_{1,x}) +3 (\Gamma_1 u_{1,t}+\Gamma_2 u_{1,y}) + \Gamma_4 +\Gamma_8 \psi_z^2}{6 (\Gamma_1 \psi_t+\Gamma_2 \psi_y)},\quad \\
	&&u_3=\Big(3 \Gamma_1 (u_{1,xt}+\psi_t u_{1,xx}- u_{2,t}+ \psi_{t}u_{2,x})+3 \Gamma_2 (u_{1,xy}+ \psi_y u_{1,xx}- u_{2,y}+ \psi_y u_{2,x}) +\Gamma_3 \psi_{yt}\nonumber\\  
	&&\qquad\qquad +\Gamma_7 \psi_{yy}+\Gamma_8 \psi_{zz}\Big)\Big/{12 ( \Gamma _1 \psi _t+\Gamma _2 \psi _y)}. 
	\eea \label{aan-eqn} \ees 
	Equations (\ref{aan-eqn}) confirm that $u_2$ and $u_3$ are not arbitrary as expected. 
	Next, from the coefficient of $\phi^{-1}$, which corresponds to the resonance $j=4$, we found that the resultant expression vanishes and leaves $u_4(x,y,z,t)$ to be arbitrary as required. Furthermore, from the coefficient of $\phi ^{0}$ (constant coefficient for the resonance $j=5$) we obtain the following expression for $u_5$: 
	\begin{align}
	u_5=& \nonumber \frac{1} {24 ( \Gamma _1 \psi _t+\Gamma _2 \psi _y)}\Big((\Gamma_6 +\Gamma_3 \psi_y) u_{2,t}+2 \Gamma_8 \psi_z u_{2,z}+\Gamma_8 u_{1,zz}+(\Gamma_5 +\Gamma_3 \psi_{t}+1) u_{2,y} +2 \Gamma_7 \psi_y\\  \nonumber
	+&\Gamma_3 u_{1,yt}+\Gamma_7 u_{1,yy}+3 \Gamma_1 u_{2,t} u_{1,x}+3 \Gamma_2 u_{2,y} u_{1,x}+2 u_3(\Gamma_4+\Gamma_8 \psi_z^2+\Gamma_7 \psi_y^2+3 \Gamma_1 u_{1,t}+3 \Gamma_2 u_{1,y}\\ \nonumber
	+&\psi_t (\Gamma_6+6 \Gamma_1 u_{2}+\Gamma_3 \psi_y+3\Gamma_1 u_{1,x})+\psi_x(\Gamma_5+6 \Gamma_2 u_{2}+3 \Gamma_2  u_{1,x}))+(2 \Gamma_4 +\Gamma_6 \psi_t +\Gamma_5 \psi_y \\ \nonumber
	+&6 \Gamma_1 u_{1,t} +6 \Gamma_2 u_{1,y}+3 \Gamma_1 \psi_{t} u_{1,x} +3 \Gamma_2 \psi_t u_{1,x})u_{2,x}-24 (\Gamma_1 \psi_t + \Gamma_2 \psi_y) u_{4,x}+(\Gamma_5 +\Gamma_6 +3\Gamma_1 u_{1,x}\\ \nonumber
	+& 3\Gamma_2 u_{1,x}) u_{1,xt}+\Gamma_4 u_{1,xx}+\ \Gamma_1 u_{1,t} u_{1,xx}+3 \Gamma_2 u_{1,y} u_{1, xx}+u_2(\Gamma_8 \psi_{zz}+\Gamma_3 \psi_{yt}+\Gamma_7 \psi_{yy} \\ \nonumber
	+ & 3(\Gamma_1 u_{2,t}+\Gamma_1 u_{1,xt}+\psi_t(3 u_{2,x}+u_{1,xx}))+\Gamma_2(u_{2,y}+u_{1,xy}+\psi_y(3 u_{2,x}+u_{1,xx})))\\
	+& 3\Gamma_1 u_{2,xxt}+3\Gamma_2 u_{2,xxy}+ (\Gamma_1 \psi_t +\Gamma_2 \psi_y) u_{2,xxx}+\Gamma_1 u_{1,xxxt}+\Gamma_2 u_{1,xxxy}\Big). \label{u5-eqn}
	\end{align}
	The above equation (\ref{u5-eqn}) confirms the non-arbitrariness of $u_5$ as we do not have any resonances at $j=5$. 
	Finally, collecting the coefficient of $\phi$ (resulting for the resonance $j=6$), the following expression is obtained:
	\bea
	&&24 u_3 ^2 (\Gamma_1 \psi _t + \Gamma_2 \psi _y )+ 2( \Gamma_6 + 3 \Gamma_1 u_2 + \Gamma_3 \psi_{y}) u_{3,t}+12 \Gamma_1 u_{5,t}+4 \Gamma_8 \psi_z u_{3,z}+\Gamma_8 u_{2,zz} \nonumber \\
	&&+2( \Gamma_5+3 \Gamma_2 u_{2} + \Gamma_3 \psi_t +2 \Gamma_7 \psi_y) u_{3,y}+12 \Gamma_2 u_{5,y}+\Gamma_3 u_{2,yt}+\Gamma_7 u_{2,yy}+6 (\Gamma_1 u_{3,t} + \Gamma_2 u_{3,y}) u_{1,x}\nonumber \\
	&&+6 u_{4}(\Gamma_4+\Gamma_8 \psi_z^2+\Gamma_7 \psi_y^2+3 \Gamma_1 u_{1,t}+3\Gamma_2 u_{1,y}+\psi_t(\Gamma_6+6 \Gamma_1 u_2+\Gamma_3 \psi_y+3 \Gamma_1 u_{1,x})\nonumber \\
	&&+\psi_y(\Gamma_5+6\Gamma_2 u_2+3 \Gamma_2 u_{1,x}))+3u_{2,x}(3 \Gamma_1 u_{2,t} +3 \Gamma_2 u_{2,y} +(\Gamma_1 \psi_t +\Gamma_2 \psi_y) u_{2,x}) \nonumber \\
	&&+(4\Gamma_4+2\Gamma_5 \psi_y +2 \Gamma_6 \psi_t +6 \Gamma_1 (3 u_{2} \psi_t+2 u_{1,t}+  \psi_t u_{1,x})+6 \Gamma_2(3 u_2 \psi_y  +2 u_{1,y} + \psi_y u_{1,x})) u_{3,x}\nonumber \\
	&&+12 (\Gamma_1 \psi_t +\Gamma_2 \psi_y) u_{5,x}+3\Gamma_1 u_{2,x} u_{1,xt}+(\Gamma_6 +3\Gamma_1 u_2 +3 \Gamma_1 u_{1,x})u_{2,xt} +12\Gamma_1 u_{4,xt}+3\Gamma_2 u_{2,x} u_{1,xy}\nonumber \\
	&&+\Gamma_5 u_{2,xy}+3 \Gamma_2 (u_{2} +u_{1,x}) u_{2,xy}+12 \Gamma_2 u_{4,xy}+3 (\Gamma_1 u_{2,t} +\Gamma_2 u_{2,y})u_{1,xx}\nonumber \\
	&&+2 u_3(\Gamma_8 \psi_{zz}+\Gamma_3 \psi_{yt}+\Gamma_7 \psi_{xx}+3\Gamma_1 (2 u_{2,t}+u_{1,xt}+\psi_t(4 u_{2,x}+u_{1,xx})\nonumber \\
	&&+\Gamma_2(2 u_{2,y}+u_{1,xy}+\psi_y(4 u_{2,x}+u_{1,xx}))))+(3 \Gamma_1 (u_2 \psi_t +u_{1,t})+3 \Gamma_2 (u_2 \psi_y +u_{1,y})+\Gamma_4)u_{2,xx} \nonumber \\
	&&+12 (\Gamma_1 \psi_t +\Gamma_2 \psi_y) u_{4,xx}+6 \Gamma_1 u_{3,xxt}+6 \Gamma_2 u_{3,xxy}+2 (\Gamma_1 \psi_t + \Gamma_2 \psi_y) u_{3,xxx}+\Gamma_1 u_{2,xxxt}+\Gamma_2 u_{2,xxxy}=0.~~~~
	\label{peq6} \eea 
	We can understand that the above equation (\ref{peq6}) becomes more complex after substituting $u_2, u_3,$ and $u_5$, which is highly nonlinear with the model parameters $\Gamma_i$. We find that this equation (\ref{peq6}) can not be satisfied for any choices of non-vanishing parameters $\Gamma_i \neq 0, i=1,2,3,\dots,8$, which is a required condition and its failure confirms that our considered model does not passes the Painlev\'e test. Hence, we can conclude that the general (3+1)D HSI equation (\ref{eq13}) is ``non-integrable in the Painlev\'e sense" because of the non-availability of required number of arbitrary functions (mainly, at the resonance $j=6$).
	
	Here we look for possible choices of $\Gamma_i$ parameters using the compatibility condition (\ref{peq6}) satisfying which the model (\ref{eq13}) can turn out to be integrable. Note that this is not an easier task due to the highly complex nature of equation  (\ref{peq6}). So, we have evaluated all versions mentioned below Eq. (\ref{eq13}) in the Introduction and found that only for two cases (choice (ii) and (iii)) Eq. (\ref{peq6}) is satisfied and hints the integrability. We have also performed the analysis thoroughly for these two choices and identified they Painlev\'e integrability, while all other cases/models are non-integrable in the Painlev\'e sense. Note that for the above two integrable choices of $\Gamma_i$ the (3+1)D equation (\ref{eq13}) reduces to (2+1)D counterparts. Interestingly, we find a new set of $\Gamma_i$ parameters as $\Gamma_1=\Gamma_3=\Gamma_6=a_1,  \Gamma_2=\Gamma_5=\Gamma_7=a_2, \Gamma_8=0$, $\Gamma_4=a_3$ the model (\ref{eq13}) becomes the following new Painlev\'e integrable version: 
	\bea a_1 [3(u_x u_t)_x+u_{xxxt}+ u_{xt}+u_{yt}]+a_2[3(u_x u_y)_x+u_{xxxy}+u_{xy}+u_{yy}]+ a_3 u_{xx} =0, \label{eq130}
	\eea 
	where $a_1,a_2$ and $a_3$ are arbitrary. Again, for confirmation, we have performed the Painlev\'e analysis for (\ref{eq130}) from the beginning (leading order analysis, resonances and arbitrary analysis) and found that it passes the Painlev\'e test without any difficulty. The equation (\ref{eq130}) describing shallow waters in (2+1)D needs a separate investigation as no reports are available so far to the best of our knowledge. Work is in progress along this direction to explore the dynamics of solitons, periodic waves, breathers, and rogue waves corresponding to integrable model (\ref{eq130}) and the results will be reported elsewhere. 
	Though the general model (\ref{eq13})  is non-integrable, we proceed to check the possibility for constructing higher-order rogue wave solutions and analyse their dynamics.

	\section{Methodology to Construct Higher-Order Rogue Wave Solutions} 
	As mentioned in the introduction, our objective is to construct rogue wave solution of (3+1)D HSI equation (\ref{eq13}) by using the Hirota's bilinear form and generalized polynomial approach. The bilinear form can be possible for most of the integrable nonlinear models and rarely a few non-integrable equations too. It is shown that the considered methodology to construct higher-order rogue waves was proven to be effective with higher-dimensional nonlinear models too  \cite{hi,pa,ss,zh,yang21}. The primary step in the construction of solutions using Hirota bilinear forms is the appropriate test functions. For $N$-soliton solutions, the exponential test functions are being used widely in the literature. Additionally, other localized structures such as breather, lump are extracted either from the $N$-soliton or through appropriate test function. Interaction solutions are primarily obtained through the combination of two different test functions. Generally, there are three approaches to choose test functions and so infinite solutions can be extracted. The $N$-soliton solutions using exponential test functions as mentioned above, and quasi-periodic (not localized) solutions through Riemann-theta functions are being used along with Hirota bilinear forms \cite{bk1}. The general rogue wave solutions are constructed by using another classical approach known as KP hierarchy reduction method. This classical approach is widely used in literature, comprising Gram determinants/Schur polynomials and Hirota bilinear forms \cite{otha1, otha2}. 
	
	The method adopted here is a recursive approach to construct higher-order rogue wave solutions of a given nonlinear soliton equations and it utilizes Hirota bilinear form(s) along with general class of polynomials.
	The approach is inspired by certain associated reports on the rogue wave solutions of Boussinesq and KP type equations \cite{pa,ss,zh}.  This method is comparatively new compared to the above three mentioned approaches for constructing solutions comprising the Hirota bilinear forms and generalized test functions. Recently,  Clarkson and Dowie solves the (1+1)D  Boussinesq equation using the considered approach \cite{pa}, also it is reported recently that the zeros of these polynomials have interesting patterns and the integral of the solutions representations are also well behaved \cite{pa, pa2}. 
	The considered approach is further extended to higher dimensional soliton equation in (2+1)D, (3+1)D, (4+1)D, and its nonlocal Alice Bob equivalent models \cite{ss, aaa1, aaa2, aaa3}. In \cite{aaa3}, it was predicted that the method is applicable for soliton equation whose bilinear form is free from mixed Hirota $D$-operators, but recently this approach is used for nonlinear models whose consist of mixed Hirota $D$-operators \cite{aaa4}. Also, it was mentioned in \cite{ss} that this method's utility is nothing to do with Painlev\'e integrability. This technique is recently implemented for complex nonlinear wave equation even without transforming into Hirota bilinear form \cite{aaa5}. 
	This motivates us to study higher-dimensional models using the considered approach. The main steps of the considered methodology is given below.
	
	Consider any (3+1)D nonlinear partial differential equation of the following form:
	\begin{equation}
	\mathcal{F}(u, u_x, u_y, u_z, u_t, u_{xx}, u_{yy}, u_{xy}, u_{zz}, u_{tt},u_{xt},u_{yt}... )=0. \label{sec31}
	\end{equation}
	\subsubsection*{Step 1:}
	To find rogue wave solutions, first we convert the above nonlinear (3+1)-dimensional model (\ref{sec31}) into a simpler (1+1)-dimensional nonlinear equation through a suitable transformation. For example, the transformation $\delta=x+\delta_1 y+ \delta_2 t$ will result eq. (\ref{sec31}) to the form given below.
	\begin{equation}
	\mathcal{G}(u, u_\delta, u_{\delta \delta}, u_{\delta \delta \delta}, \dots, u_z, u_{zz})=0. \label{sec33}
	\end{equation}
	
	\subsubsection*{Step 2:}
	The next step is to obtain bilinear form of the dimension-reduced nonlinear equation (\ref{sec33}). For this purpose, a suitable bilinearizing transformation is identified from the leading order analysis of the Painlev\'e test. It should be mentioned that there a couple of bilinearizing transformations, namely logarithmic and rational functions, are widely used to find solutions of a number of nonlinear equations. Let us take $\mathcal{H}$ is the bilinearizing transformation with respect the function $\mathcal{R}(\delta, z)$ as given below. 
	\begin{equation}
	u(\delta, z)=\mathcal{H}(\mathcal{R}(\delta, z)). \label{sec32}
	\end{equation}
	
	\subsubsection*{Step 3:}
	Through the bilinearizing transformation (\ref{sec32}), one can deduce  bilinear form of the (1+1)-dimensional equation (\ref{sec33}) utilizing the Hirota derivatives \cite{hi}.
	\begin{equation}
	\mathcal{N}(D_\delta, D_z; \mathcal{R})=0, \label{sec35}
	\end{equation}
	where $D$ represents the Hirota bilinear operator \cite{hi} and it can be defined as follows consisting of two functions.
	\begin{equation}
	D_\delta^p D_z^q (f \cdot g)=(\partial_\delta-\partial_\delta')^p (\partial_z-\partial_z')^q f \cdot g|_{\delta'=\delta,z'=z}.  \label{sec36} 
	\end{equation}
	
	\subsubsection*{Step 4:}
	Now, we consider the following generalized polynomial test function for $\mathcal{R}$:
	\bes\begin{equation}
	\mathcal{R}  = \mathcal{R}_{r+1} (\delta,z,\lambda, \mu ) = \mathcal{R}_{r+1} (\delta, z) + 2 \lambda z \mathcal{P}_r (\delta, z) + 2 \mu \delta \mathcal{Q}_r (\delta, z) + (\lambda^2 +\mu ^2 ) \mathcal{R} _{r-1},
	\end{equation}
	with 
	\bea
	\mathcal{R}_{r} (\delta, z) & = \ds \sum _{k=0} ^{l(l+1)/2} \sum _{j=0}^{k} \chi_{l(l+1)-2k,2j} z ^{2j} \delta ^{l(l+1)-2k } , \\
	\mathcal{P}_r (\delta, z) & =  \ds\sum _{k=0} ^{l(l+1)/2} \sum _{j=0}^{k} \phi_{l(l+1)-2k,2j} z ^{2j} \delta ^{l(l+1)-2k } , \\
	\mathcal{Q}_r(\delta, z)& = \ds\sum _{k=0} ^{l(l+1)/2} \sum _{j=0}^{k} \psi_{l(l+1)-2k,2j} z ^{2j} \delta ^{l(l+1)-2k}.
	\eea \label{eq27}\ees 
	Here $\lambda, \mu , \chi_{a,b}, \phi_{a,b}$ and $\psi_{a,b} (a,b=0,2,4, \ldots , l(l+1))$ are real parameters, and $\mathcal{R}_{-1}=\mathcal{P}_0=\mathcal{Q}_0=0$.
	
	\subsubsection*{Step 5:} 
	The final step in the process of constructing solution is to adopt the above generalized polynomial (\ref{eq27}) up to the required order for the bilinear function and on substituting it into the bilinear equation(s) (\ref{sec35}), one can arrive at a set of equations arising as the coefficients of different powers of $z^c \delta^e$. Solving those equations recursively will provide exact form of the parameters $\chi_{a,b}$, $\phi_{a,b}$ and $\psi_{a,b}$, where $a,b=0,2,4, \ldots , l(l+1)$, from that the required solution of the original nonlinear equation can be obtained. 
	
	\section{Rogue Wave Solutions}
	In this section, we extract rogue wave solutions by using the methodology given in the previous section 3. Mathematically, rogue waves are rational solutions of nonlinear models. To do so, first we use the transformation $\delta(x,y,t)=x+\delta_1 y+\delta_2 t$, for dimensional reduction of Eq. (\ref{eq13}) which will result into $u (x,y,z,t)\rightarrow u (\delta,z)$ and can be written in the following form:
	\begin{equation}
	{ (\Gamma_1 \delta_2+\Gamma_2 \delta_1)  u_{\delta \delta \delta \delta} +6 (\Gamma_1 \delta_2+\Gamma_2 \delta_1) u_\delta u_{\delta \delta}+(\Gamma_4+\Gamma_5 \delta_1+\Gamma_7 \delta_1^2+\Gamma_6 \delta_2+\Gamma_3 \delta_1\delta_2) u_{\delta \delta}+ \Gamma_8 u_{zz}=0.} \label{eq31}
	\end{equation}   
	Though our considered model (\ref{eq13}) is non-integrable in the Painlev\'e sense, we adopt the following logarithmic transformation to bilinearize the above nonlinear equation (\ref{eq31}):
	\bea u(\delta,z)=u_0+2 \frac{\partial}{\partial \delta}[ ln \mathcal{R} (\delta,z)].  \label{bieq}\eea 
	From Eqs. (\ref{eq31}) and (\ref{bieq}), we obtain the following Hirota bilinear form:
	\begin{equation}
	(\Gamma_1 \delta_2+\Gamma_2 \delta_1) D_\delta^4+(\Gamma_4+\Gamma_5 \delta_1+\Gamma_7 \delta_1^2+\Gamma_6 \delta_2+\Gamma_3 \delta_1\delta_2) D_\delta^2  +\Gamma_8 D_z^2)\mathcal{R}.\mathcal{R}=0, \label{eq32a} 
	\end{equation}
	where $D$ is Hirota bilinear operator \cite{hi} as defined in (\ref{sec36}). In the above bilinear form (\ref{eq32a}), the model parameters $\Gamma_1$ and $\Gamma_2$ are responsible for the fourth-order Hirota operation in spatio-temporal dimension ($D_\delta^4$), while $\Gamma_3,\Gamma_4, \Gamma_5, \Gamma_6$ and $\Gamma_7$ are responsible for its second-order effects ($D_\delta^2$) and the second-order spatial operation ($D_z^2 $) is controlled by the model parameter $\Gamma_8$. To simplify the computational complexity, without loss of generality, we consider those three different parts by assuming the choice $\Gamma_1=\Gamma_2=\alpha$,  $\Gamma_3=\Gamma_4=\Gamma_5=\Gamma_6=\Gamma_7=\beta$, and $\Gamma_8=\gamma$, which 
	reduces equation (\ref{eq32a}) to a simple version of the bilinear form as given below.
	\begin{equation}
	\alpha( \delta_2+ \delta_1) D_\delta^4+\beta(1+\delta_1+ \delta_1^2+ \delta_2+\delta_1\delta_2) D_\delta^2  +\gamma D_z^2)\mathcal{R}.\mathcal{R}=0.\label{eq32} 
	\end{equation}
	Thus the corresponding model equation pertinent to the above bilinear representation (\ref{eq32}) takes the following form, from (\ref{eq13}):
	\begin{equation}
	\alpha( 3(u_x u_t)_x+u_{xxxt}+3(u_x u_y)_x+u_{xxxy})+\beta( u_{yt}+  u_{xx}+  u_{xy}+  u_{xt}+  u_{yy})+\gamma u_{zz}=0. \label{eq2a}
	\end{equation} 
	
	Henceforth, our attention is limited to the above version of (3+1)-D HSI equation (\ref{eq2a}). In the following part, we shall construct explicit first-, second-, and third-order rogue wave solutions (of order one, two and three) by considering the bilinear transformation (\ref{bieq}) and the polynomial functions (\ref{eq27}) up to a required order. Further, we shall explore the dynamics of those rogue waves in details with appropriate analysis and necessary graphical demonstrations.
	
	\subsection{Rogue wave of order one}
	To construct a first-order rogue wave solution, we consider the lowest order parameter in the series expansion (\ref{eq27}) which is nothing but $r=0$. This results into the initial form of the function $\mathcal{R}$ as 
	\begin{equation}
	\mathcal{R}=\mathcal{R}_1(\delta , z ) =  \chi  _{0,0} + \chi _{2,0} \delta ^2 + \chi _{0,2} z^2, \label{eq1} 
	\end{equation}
	where $\chi _{0,0}$, $\chi _{0,2}$, and $\chi _{2,0}$ are arbitrary parameters. Without loss of generality, we take a choice $\chi _{2,0} =1$ and substituting above equation (\ref{eq1}) into bilinear form (\ref{eq32}), we obtain a polynomial equation in $z$ and $\delta$ as follows.
	\begin{equation}
	6 \alpha  K -\beta L(\delta^2-\chi_{0,0}-z^2 \chi_{0,2})+ \gamma \chi_{0,2}  (\delta^2+\chi_{0,0}-z^2 \chi_{0,2})=0, \label{eq312}
	\end{equation}
	where $K=(\delta_1 + \delta _2)$ and $L=(1+\delta_1+\delta_2+\delta_1^2+\delta_1\delta_2)$ taken for simplicity in the notation. On collecting the coefficients of $z^2$, $\delta^2$ and constant terms, we get the following equations:
	\bes\bea
	&\beta \chi _{0,2} L - \gamma \chi _{0,2}^2  =0,& \\
	&6 \alpha  K  + \beta \chi _{0,0} L+ \gamma \chi _{0,0}\chi _{0,2} =0.& 
	\eea \label{eq2}\ees
	On solving the above couple of equations in a straightforward way, one can easily gets the form of remaining two parameters in terms of other arbitrary constants
	\begin{align}
	\chi _{0,0} & = {- 3 \alpha K }\big/{\beta L}, \quad 
	\chi _{0,2}  = {\beta L}/{\gamma }. \label{eq3}
	\end{align}
	Thus, the explicit form of $\mathcal{R}_1$ becomes
	\begin{equation}
	\mathcal{R}_1(\delta , z ) = \beta \gamma L\delta^2 - {3 \alpha \gamma  K }+ {\beta^2 L^2 z^2}. \label{eq4}
	\end{equation}
	Finally, the resulting rogue wave solution of order one to equation (\ref{eq2a}) can be obtained by using the above $\mathcal{R}_1$ (\ref{eq4}) and the bilinear transformation  (\ref{bieq}) as given below in a simplified form.
	\begin{equation}
	u(x,y,z,t)  = u_0+\dfrac{4\beta \gamma L (x + \delta _1 y + \delta_2 t)}{\beta \gamma L (x+ \delta _1 y + \delta _2 t )^2 + {\beta^2 L^2 z^2} - {3 \alpha \gamma K  }}, \label{eq5} 
	\end{equation}
	with $K=(\delta_1 + \delta _2 )$ and $L=(1+\delta_1 + \delta_1 ^2 + \delta _ 2 + \delta _1 \delta _2)$. Moreover, there exists a $\lim_{|x| \to \infty}u_1(x,y,z,t)=u_0$, $\lim_{|y| \to \infty}u_1(x,y,z,t)=u_0$ and $\lim_{|z| \to \infty}u_1(x,y,z,t)=u_0$. This represents that the solution decays to the background $u_0$ along all the spatial directions, which can be either zero or non-zero. From (\ref{eq32}) , It is clear that the background $u_0$ (arbitrary parameter) considered in the logarithmic transformation (\ref{bieq}) does not emerge in bilinear form (\ref{eq32}), hence the background of rogue waves can be either zero or non-zero. It is also observed that from another recent work \cite{ss} that rogue wave appears under the constraint condition depending on arbitrary background parameter $u_0$.
	
\begin{figure}[h]
		\centering\includegraphics[width=0.95\linewidth]{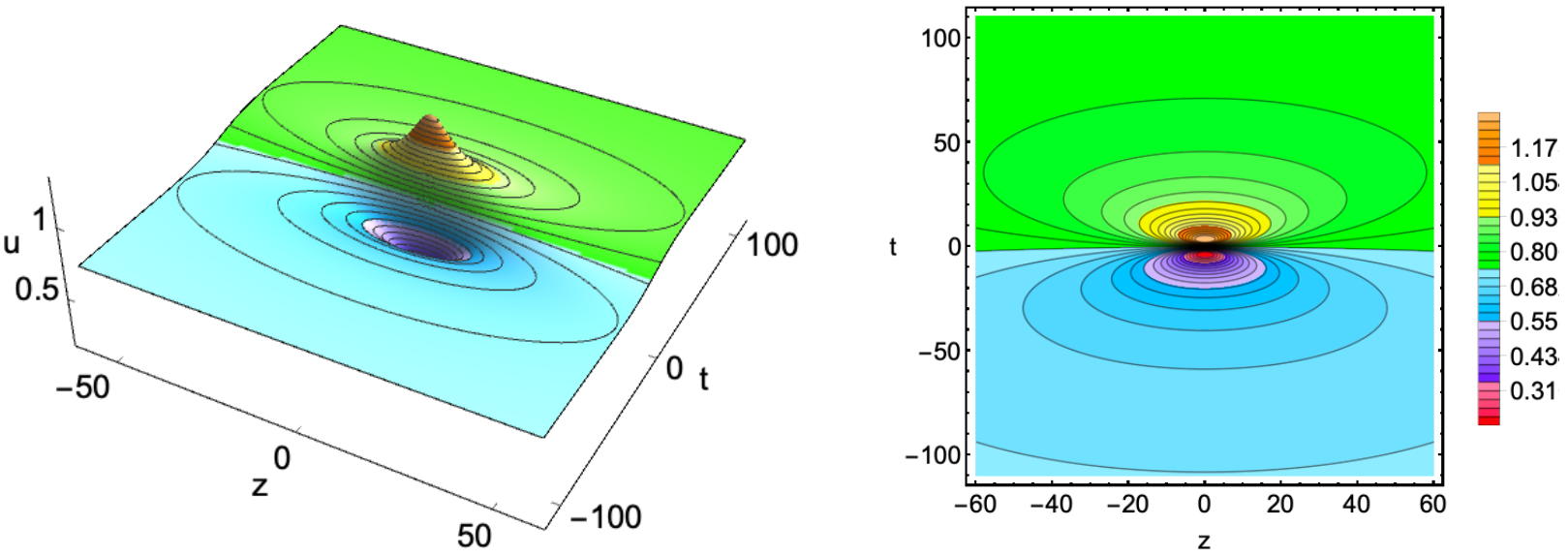}
		\caption{Nature of the first-order rogue wave (\ref{eq5}) consisting of a doubly-localized up-down single-peak in $z-t$ for $y=0.1$ and $x=0.3$ with the choice of other parameters as $u_0=0.75$, $\alpha=1.5$, $\beta=0.15$, $\gamma=0.5$, $\delta_1=-1.5$, and $\delta_2=1.0$.}
		\label{fig-first}
	\end{figure}

	The above first-order rogue wave solution (\ref{eq5}) consists of six arbitrary parameters $u_0$, $\alpha$, $\beta$, $\gamma$, $\delta_1$, and $\delta_2$. A necessary condition that has to be satisfied for constructing a non-singular regular structure is $L\neq 0$, which is driven by the parameters $\delta_1$ and $\delta_2$ that play significant role in the dynamics of resulting rogue waves. Through a careful analysis on the evolution of solution (\ref{eq5}), we find that it results a doubly-localized rogue wave of a special type with one peak upward and another downward-dip on a background $u_0$ in the $z-t$ plane as shown in Fig. \ref{fig-first}. This is quite different from the standard rogue wave having one-central peak and a dip on its either side \cite{aa8,aa7,aa9,aa10,aa11,aa12,aa13,aa14}. It should also be noted that the above two mentioned localizations strongly depends upon the logarithmic transformation. 
	
\begin{figure}[h]
		\centering\includegraphics[width=0.975\linewidth]{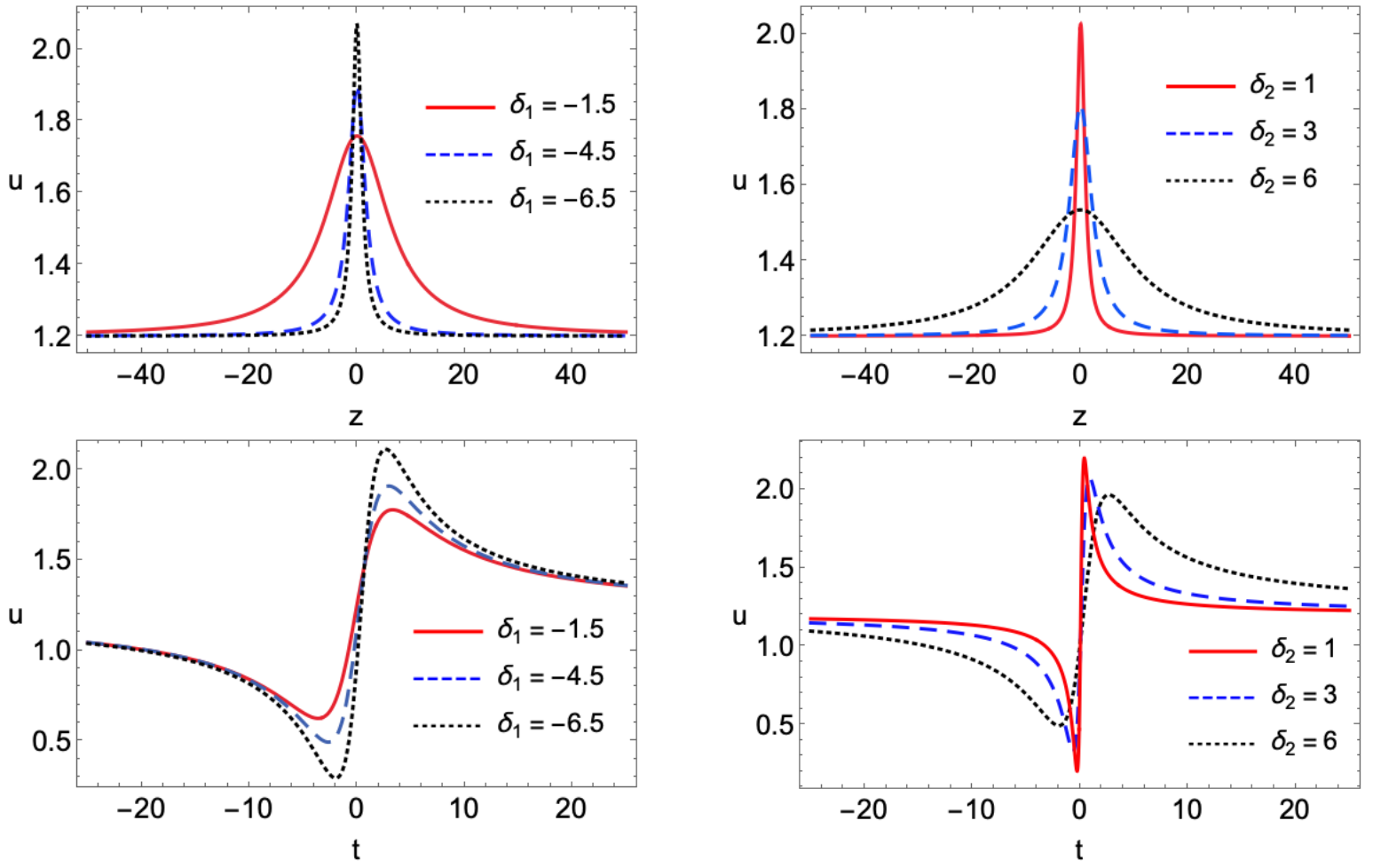} 
		\caption{Role of the arbitrary parameters $\delta_1$ and $\delta_2$ in controlling the peak-amplitude and width of the first-order rogue wave (\ref{eq5}) along spatial direction $z$ (top panels) for $t=0.1$ and along time $t$ (bottom panels) for $z=0.1$ with $x=0.3$, $y=0.1$ and the other parameter values given in Fig. \ref{fig-first}.}
		\label{fig-first2}
	\end{figure}

	It is important to understand the impact of arbitrary parameters in determining the peak/dip amplitude, length, and width of the rogue wave. One can easily identify that the role of $u_0$ is to serve the background amplitude without affecting the pattern formation and the nature of rogue wave in any way. The parameters $\delta_1$ and $\delta_2$ play crucial role in controlling amplitude, length, and width of the rogue wave. To be specific, increasing $\delta_1$ increases the amplitude with compression of width along $z$, while it does not alters the length of the rogue wave along $t$. On the other hand, $\delta_2$ decreases the amplitude of rogue wave along with commensurate widening of the wave structure along both $z$ and $t$ axes as shown in Fig. \ref{fig-first2}. Further, the system parameters $\alpha$, $\beta$, $\gamma$ appearing in the solution also manipulates the rogue wave behaviour. Mainly, there occurs a change (increase) in the width of the doubly-localized rogue waves induced by the increase in $\gamma$, while the parameters $\alpha$ and $\beta$ reduces their width on either side along $z$ and $t$ and we have restricted the graphical depiction considering the length of the article. 
	
	\begin{figure}[h]
		\centering\includegraphics[width=0.95\columnwidth]{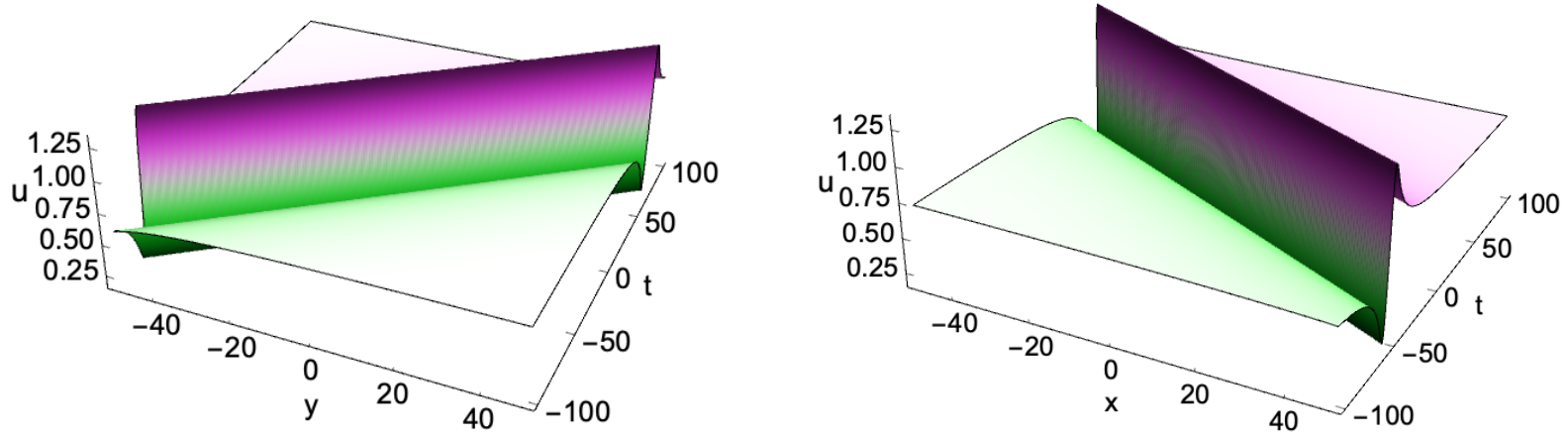}
		\caption{Nature of the first-order line rogue wave (\ref{eq5}) in $y-t$ for $x=0.3$ and in $x-t$ for $y=0.3$ with $z=0.1$ and by keeping the other parameters same as in Fig. \ref{fig-first}.}
		\label{fig-first3}
	\end{figure}
	
	The solution (\ref{eq5}) also exhibit a different type of localized wave profiles which are nothing but line rogue waves along the planes $x-t$ and $y-t$ with single up and down steady amplitudes. This is also of different from that of rational solitons which usually posses an amplitude dip on both sides of a stable peak structure. Here the line rogue wave exhibits only single peak and dip in its amplitude above and below the constant background, respectively. Further, the background is also of little asymmetric in amplitude resembling line rogue wave on a tiny kink background. For illustrative purpose, we have shown a first-order line rogue wave of the generalized HSI model in Fig. \ref{fig-first3}. As discussed in the case of doubly-localized rogue waves, the manifestation and control of these line rogue waves can also be modified by suitably tuned arbitrary parameters. 
	
	\subsection{Rogue wave of order two}
	We can deduce rogue wave solution of order two by considering the order parameter $r=1$ in the generalized polynomial function (\ref{eq27}) and it results into the following form: 
	\begin{align}
	\mathcal{R}  = \mathcal{R}_{2} (\delta,z,\lambda, \mu ) & = \chi _{0,0} + \chi _{0,2} z^2 + \chi _{0,4} z ^4  +\chi _{0,6} z^6 + ( \chi _{2,0} + \chi _{2,2}z^2 + \chi _{2,4} z^4 ) \delta ^2 + (\chi _{4,0} + \chi _{4,2} z^2 ) \delta^4 \notag \\
	& + \delta ^6  + 2 \lambda z ( \phi _{0,0}  + \phi _{0,2} z^2 + \phi _{2,0} \delta ^2 ) + 2 \mu \delta ( \psi _{0,0} + \psi _{0,2} z^2  + \psi _{2,0} \delta ^2 ) + \lambda^2 + \mu ^2. \label{eq6}
	\end{align}
	Substituting the above form of $\mathcal{R}$ (\ref{eq6}) into the bilinear equation (\ref{eq32}), we get a polynomial in terms of $z$ and $\delta$. Collecting the coefficient at different powers of $z$ and $\delta$, we can obtain a system of equations and upon solving them, we end up with the following parameters arising for the second-order rogue wave solution: 
	\bes
	\bea
	&&\chi _{0,0} =  \left[{16875 \alpha^3 K ^3   + \beta^2 L ^2 ( - \gamma \lambda ^2 \phi _{2,0} ^2  + 9 \beta L ( \lambda^2 - \mu ^2 ( \psi _{2,0}^2 -1)))}\right] \big/{9 \beta ^3 L^3 }, \\
	&&\chi _{0,2} = {475 \alpha ^2 K^2 }\big/{\beta \gamma L}, \quad
	\chi _{0,4}  = { -17 \alpha \beta  K L}\big/{\gamma ^2 }, \quad
	\chi _{0,6}  = {\beta ^3 L^3 }\big/{\gamma ^3 }, \\
	&&\chi _{2,0 } = {-125 \alpha ^2 K ^2 }\big/{\beta ^2 L ^2 },\quad
	\chi _{2,2}  = {-90 \alpha K}/{\gamma }, \quad
	\chi _{2,4}  = {3 \beta ^2 L^2 }\big/{\gamma^2}, \\
	&&\chi _{4,0} = {-25\alpha K}\big/{\beta L}, \quad
	\chi _{4,2}  = { 3 \beta L}\big/{\gamma }, \quad
	\phi _{0,0} =  {-5 \alpha K \phi _{2,0} }\big/{3 \beta L},\\
	&&\phi _{0,2} = {-\beta L \phi _{2,0}}/{3 \gamma}, \quad
	\psi _{0,0}  = {\alpha K \psi _{2,0} }/{\beta L},\quad
	\psi _{0,2}  = {- 3 \beta L \psi _{2,0} }/{\gamma},
	\eea \label{eq7}\ees
	where $K=(\delta _1+\delta _2)$ and $L=(1+\delta _1+\delta _2+\delta _1^{2}+\delta _1\delta _2)$. From the above explicit expression of the coefficients, we can obtain explicit form of $\mathcal{R}_{2}$ from Eq. (\ref{eq6}). Finally, by using the bilinear transformation (\ref{bieq}), the second-order rogue wave solution of the generalized (3+1)-dimensional Hirota-Satsuma-Ito equation (\ref{eq2a}) is deduced as follows.
	\bes\bea
	u(x,y,z,t)=u_0+2 [ln (\mathcal{R}_2)]_\delta \Rightarrow u_0+\dfrac{G_2(x,y,z,t)}{F_2(x,y,z,t)}, \label{eq12a}
	\eea
	where the exact expression of $G_2$ and $F_2$ takes the following form:
	\bea 
	G_2 &=&36{L}\beta \gamma (3{L}^{4}{z}^{4}\beta^{4}({x}+ \delta _1 y+\delta _2 t)- 125\ {K}^{2}\alpha^{2}\gamma ^{2}({x}+ \delta _1 y+\delta _2 t) \nonumber\\ 
	&& +3{L}^{3}{z}^{2}\beta^{3}\gamma (2({x}+ \delta _1 y+\delta _2 t)^{3}-\mu\psi_{ 2,0})- {K}{L}\alpha\beta \gamma ^{2}(50({x}+ \delta _1 y+\delta _2 t)^{3}-\mu\psi _{2,0})\\ 
	&& +{L}^{2}\beta^{2}\gamma ({x}+ \delta _1 y+\delta _2 t)  (-90{K}{z}^{2}\alpha+\gamma (2{z}\lambda\phi_{2,0}+3({x}+ \delta _1 y+\delta _2 t)(({x}+ \delta _1 y+\delta _2 t)^{3}+\mu\psi _{2,0})))),\nonumber \eea \bea
	F_2 &=& (9{L}^{6}{z}^{6}\beta^{6}-16875{K}^{3}\alpha^{3} \gamma^{3}+27{L}^{5}{z}^{4}\beta^{5}\gamma ({x}+ \delta _1 y+\delta _2 t)^{2}- 1125 {K}^{2}{L}\alpha^{2}\beta \gamma ^{3}({x}+ \delta _1 y+\delta _2 t)^{2}\nonumber\\ 
	&& +{L}^{2}\beta^{2}\gamma ^{2} (4275 {K}^{2}{z}^{2}\alpha^{2}+\gamma ^{2}\lambda^{2}\phi_{ 2,0}^{2}- 3\ {K}\alpha \gamma (75({x}+ \delta _1 y+\delta _2 t)^{4}+10{z}\lambda\phi _{2,0} \nonumber\\ 
	&& -6({x}+ \delta _1 y+\delta _2 t)\mu\psi _{2,0}))- 3{L}^{4}{z}^{2}\beta^{4}\gamma (51{K}{z}^{2}\alpha-9 {\gamma }({x}+ \delta _1 y+\delta _2 t)^{4}\nonumber\\ 
	&& +2 {\gamma }({z}\lambda\phi _{2,0}+9({x}+ \delta _1 y+\delta _2 t)\mu\psi _{2,0}))+ 9{L}^{3}\beta^{3}\gamma ^{2}(-90{K}{z}^{2}\alpha({x}+ \delta _1 y+\delta _2 t)^{2}\\ 
	&& +\gamma ({x}^{6}+6{x}^{5}(\delta _1 y+\delta _2 t)+15{x}^{4}(\delta _1 y+\delta _2 t)^{2}+  (\delta _1 y+\delta _2 t)^{6}+2{z}(\delta _1 y+\delta _2 t)^{2}\lambda\phi _{2,0}\nonumber\\ 
	&& +2(\delta _1 y+\delta _2 t)^{3}\mu\psi _{2,0}+\mu^{2}\psi _{2,0}^{2}+2{x}^{3} (10(\delta _1 y+\delta _2 t)^{3}+\mu\psi _{2,0})+{x}^{2}(15(\delta _1 y+\delta _2 t)^{4}\nonumber\\ 
	&& +2{z}\lambda\phi _{2,0}+6(\delta _1 y+\delta _2 t)\mu\psi _{2,0})+ 2{x}(\delta _1 y+\delta _2 t)(2{z}\lambda\phi _{2,0}+3(\delta _1 y+\delta _2 t)((\delta _1 y+\delta _2 t)^{3}+\mu\psi _{2,0}))))).\nonumber
	\eea \label{rogue2} \ees
	Note that $G_2(x,y,z,t)$ and $F_2(x,y,z,t)$ are polynomials of degree five and six, respectively. As we seen for the first order rogue wave solution, the present rogue wave solution of order two (\ref{rogue2}) decays to the constant background $u_0$ through all the spatial directions as $\lim_{|x| \to \infty}u_2(x,y,z,t)=u_0$, $\lim_{|y| \to \infty}u_2(x,y,z,t)=u_0$ and $\lim_{|z| \to \infty}u_2(x,y,z,t)=u_0$. 
	The above mentioned second-order rogue wave solution (\ref{rogue2}) consists of ten arbitrary parameters, namely $\alpha$, $\beta$, $\gamma$, $u_0$, $\delta_1$, $\delta_2$, $\phi_{2,0}$, $\psi_{2,0}$, $\lambda$ and $\mu$, where the first three originate from the model itself, while the remaining appear explicitly in the solution only. 
	\begin{figure}[h]
		\centering\includegraphics[width=0.85\columnwidth]{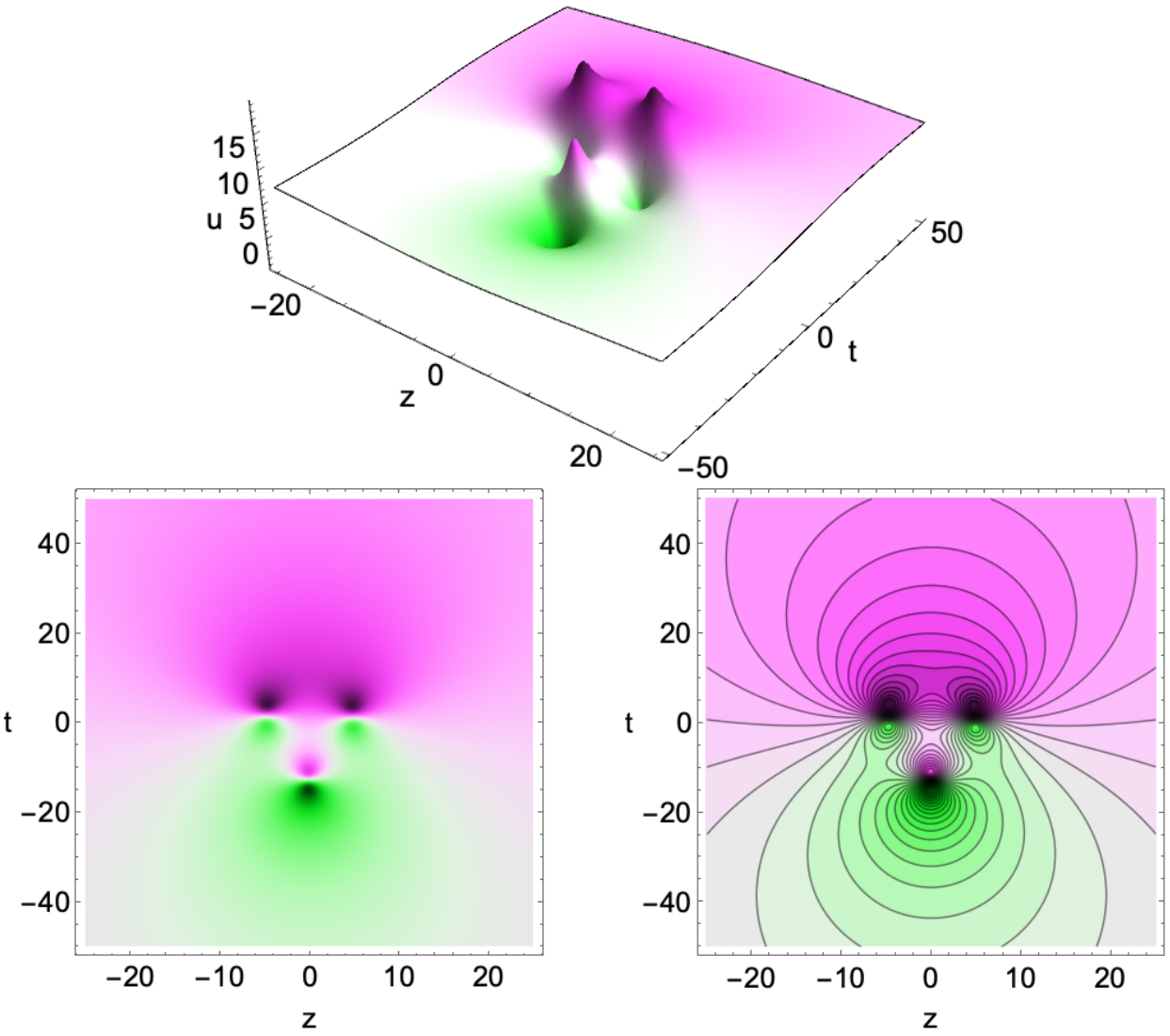}
		\caption{Nature of the second-order rogue wave exhibiting three separate doubly-localized up-down peak  (rogue wave triplet) in the triangular (face-like) structure for the choice $u_0=8.75$, $\alpha=0.02$, $\beta=-0.25$, $\gamma=-10.9$, $\delta_1=0.2$, $\delta_2=0.1$, $\phi_{2,0}=0.3$, $\psi_{2,0}=1.4$, $\lambda=0.51$ and $\mu=0.76$ with $y=0.1$ and $x=0.3$. The upper plot shows the 3D view while the bottom panel depicts the density and contour plots.}
		\label{fig-second}
	\end{figure}
	
	Similar to the first-order solution, the present second-order solution (\ref{rogue2}) also possesses both doubly-localized structures in addition to line rogue waves. First, we discuss the fascinating dynamics of doubly-localized rogue waves arising in the plane $z-t$. An interesting fact here is that there occurs more number of closely placed/packed rogue waves structures instead of single profile in the previous case. To be precise, the second-order rogue wave admits three doubly-localized structures, among which two are identical/symmetric while the third one is different and it can be referred to as rogue wave triplets. Here each of those three structures can be found to resemble the first-order rogue wave individually and each of those have a peak-amplitude and a reverse-dip from a constant background as depicted in Fig. \ref{fig-second} for an easy understanding purpose. Further, on analysing the geometrical nature the second-order rogue wave forms a triangular structure, which is also looks like a human-face patter as clearly shown in the two-dimensional density and contour plots in Fig. \ref{fig-second}. 
	\begin{figure}[h]
		\centering\includegraphics[width=01.0\columnwidth]{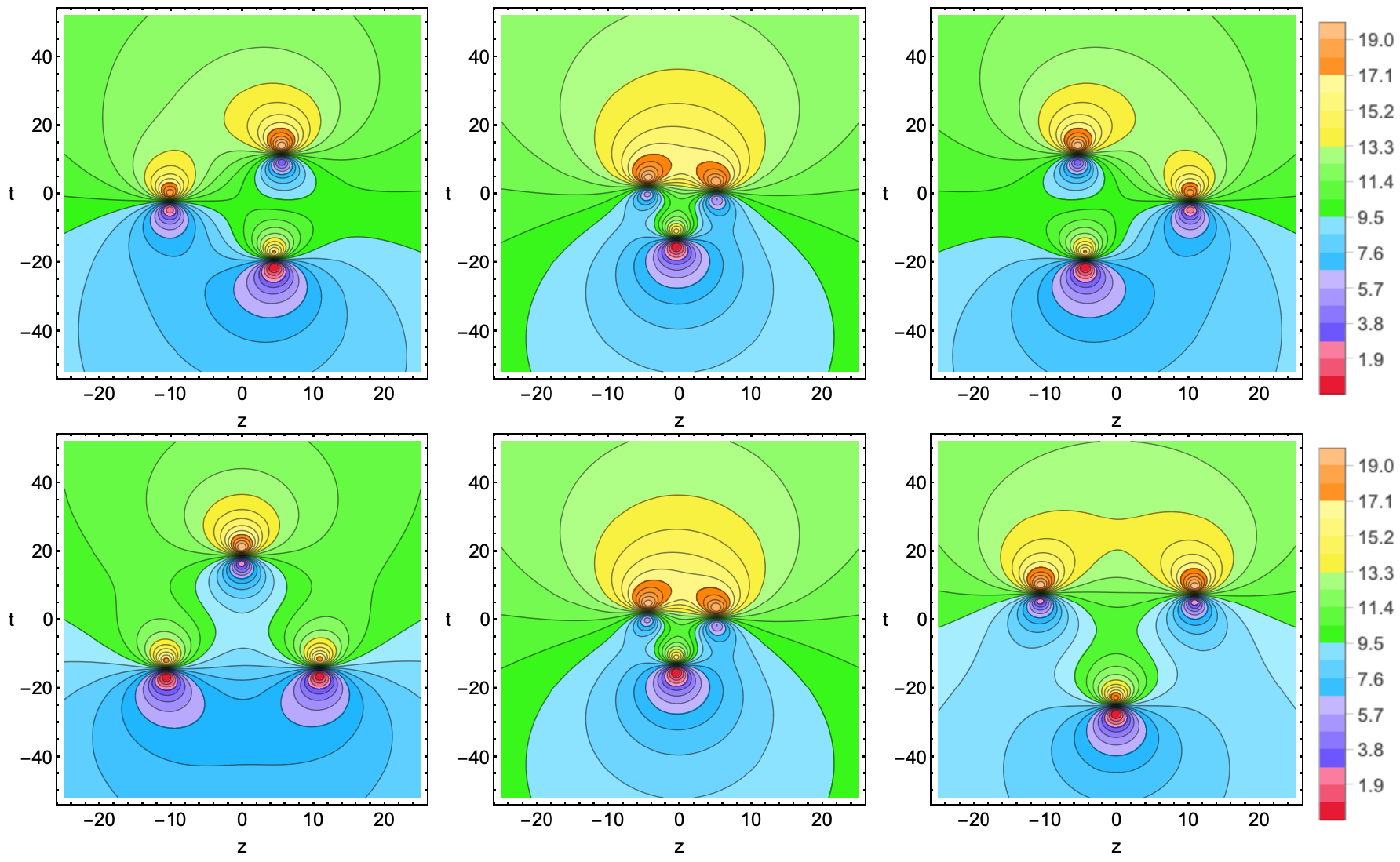}
		\caption{Effects of $\lambda$ and $\mu$ parameters in rotating the second-order rogue waves. The formation of second-order rogue wave triplets and their rotation for different choices of  (top) $\lambda=-10.51$, $\lambda=0.51$, \& $\lambda=10.51$ and (bottom) $\mu=-7.76$, $\mu=0.76$, \& $\mu=7.76$ with fixed other parameters as in Fig. \ref{fig-second}.}
		\label{fig-second2}
	\end{figure}
	
	An interesting advantage of the present second-order rogue wave solution is that the number of arbitrary parameters and each of them is contributing to engineer the rogue wave pattern as required with suitable combinations demonstrating various features. This starts from controlling the amplitude/depth, length and width of the peaks to their inter-peak spacing, rotation about different magnitudes and merger of the peaks/dips in the rogue wave group by properly chosen values of these parameters. As discussed in the first-order rogue wave, here too the $\delta_1$ and $\delta_2$ help to alter the amplitude, length and width of these individual triple structures. Further, the system parameters $\alpha$, $\beta$ and $\gamma$ induces significant change in the width of these structures. On the other hand, $u_0$ supplies the constant amplitude which can be either zero or non-zero background without altering any identities of the rogue waves. 
	
\begin{figure}[h]
		\centering\includegraphics[width=01.0\columnwidth]{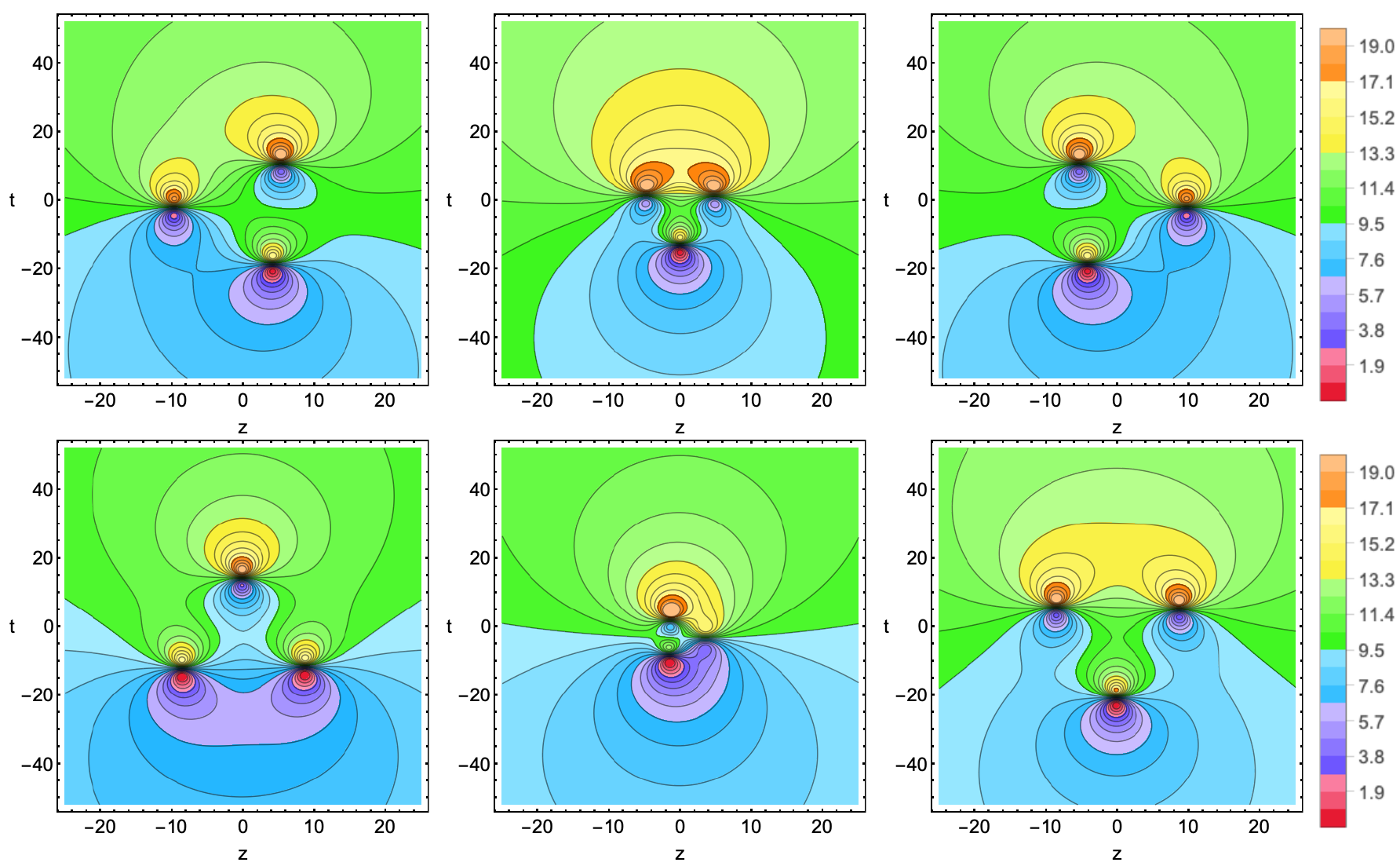}
		\caption{Effects of $\phi_{2,0}$ and $\psi_{2,0}$ parameters in rotating the second-order rogue waves. Rotation of second-order rogue wave triplets for different choices of  (top) $\phi_{2,0}=-5.03$, $\phi_{2,0}=0.03$, \& $\phi_{2,0}=5.03$ and (bottom) $\psi_{2,0}=-7.04$, $\psi_{2,0}=0.04$, \& $\psi_{2,0}=7.04$ with fixed other parameters as in Fig. \ref{fig-second}.}
		\label{fig-second3}
	\end{figure}

\begin{figure}[h]
		\centering\includegraphics[width=0.845\columnwidth]{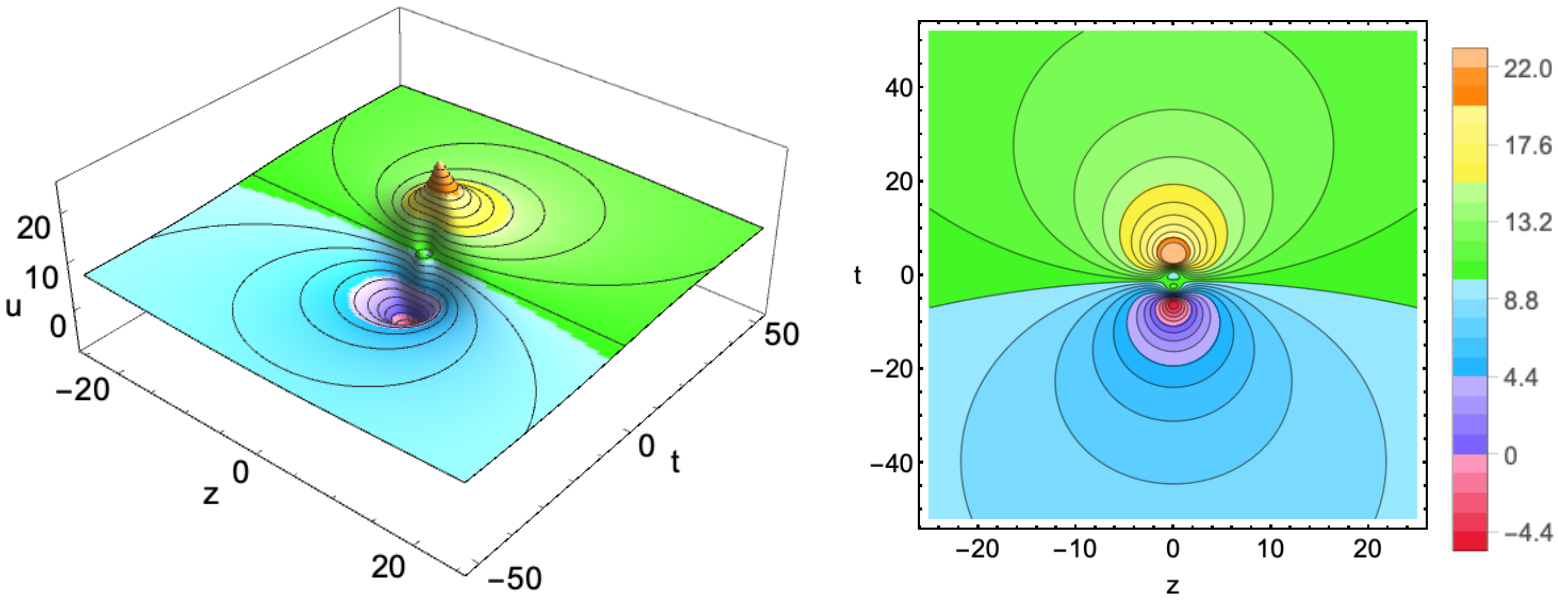}
		\caption{Simple structure of the second-order rogue waves with degenerated/merged peaks/dips resembling the first-order rogue wave form for $\lambda=\mu=\phi_{2,0}=\psi_{2,0}=0$ and other parameters are as in Fig. \ref{fig-second}.}
		\label{fig-second4}
	\end{figure}

	The role of other four arbitrary parameters $\lambda$, $\mu$, $\phi_{2,0}$, and $\psi_{2,0}$ is much more interesting that signify the structural orientation, inter-peak spacing, rotational characteristics, and merging/overlapping apart from the standard role on their amplitude, width, and length of each peak/dip in the rogue wave triplet. Smaller the magnitude of these parameters narrow closely-packed the triplets are, whereas the higher magnitudes widen them by increasing the inter-peak/dip distances too. Further, their signs determines their orientation by rotating them in opposite directions. Here the effects of $\lambda$ and $\phi_{2,0}$ looks similar, while $\mu$  and $\psi_{2,0}$ varies the triplets in the same way only with a significant change induced by $\psi_{2,0}$ which has the ability to compress the given rogue wave structure appropriately. Upon the negligible impact of these four parameters ($\lambda\approx\mu\approx\phi_{2,0}\approx\psi_{2,0}\approx0$) the present second-order rogue wave triplets straightforwardly reduces similar (but, not exactly) to that of first-order rogue wave structure. All such roles and influences of these parameters are depicted in Figs. \ref{fig-second2}, \ref{fig-second3} and \ref{fig-second4}.
	
\begin{figure}[h]
		\centering\includegraphics[width=0.9\columnwidth]{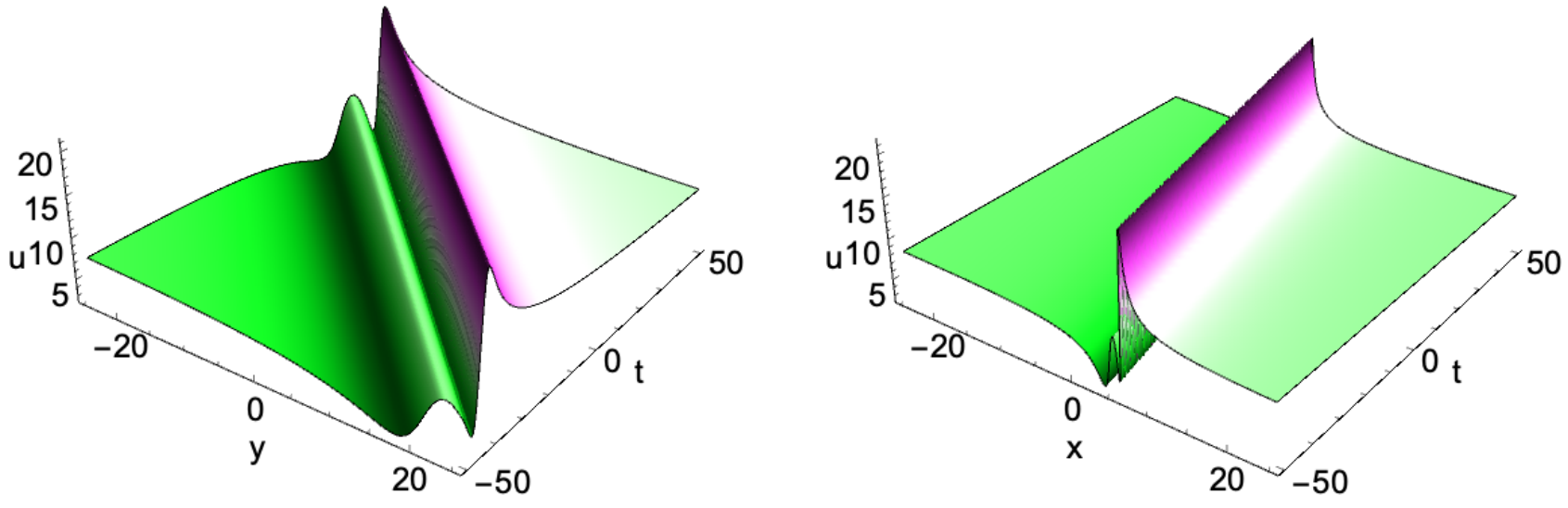}
		\caption{Second-order line rogue waves in $y-t$ possessing two peaks and dips for $z=0.1$ and $x=0.3$ (a) and in $x-t$ for $z=0.1$ and $y=0.3$ with other  parameters similar to Fig. \ref{fig-second}.}
		\label{fig-second5}
	\end{figure}

	Next, we should also consider their dynamics along the other planes such as $y-t$ and $x-t$, where the second-order rogue wave exhibits two amplitude-peaks and two dips of different heights/depths. These can be referred to as second-order rational solitons that are nothing but the overlap/superposition of first order ones. For completeness, we have illustrated such second-order rational wave type structures in Fig. \ref{fig-second5}. Further, by selecting appropriate choices of parameters, the identities like line rogue wave amplitudes, width, velocity or angle of orientation can be controlled in addition to the known background amplitude $u_0$. Moreover, it is worth to mention that the parameter $\phi_{2,0}$, $\psi_{2,0}$, $\lambda$ and $\mu$ appearing in the second order rogue wave solution that are absent in the first order rogue wave solution provide additional freedom in controlling the nature of localized wave structures. 	
	
	\subsection{Rogue wave of order three}
	To construct the rogue wave solution of order three, we put $r=2$ in the generalized function (\ref{eq27}) and the following simple polynomial test function is obtained:
	\begin{align}
	\begin{split}
	\mathcal{R}  = \mathcal{R}_{3} (\delta,z,\lambda, \mu ) &  =  
	\chi _{0,0} + \chi _{0,2} z ^2 + \chi _{0,4}z^4 + \chi _{0,6} z^6  + \chi _{0,8} z^8  + \chi _{0,10} z ^{10 } + \chi _{0,12} z^{12 } \\
	& \quad   + ( \chi _{2,0} + \chi _{2,2}z^2  +\chi _{2,4} z ^4 + \chi _{2,6} z^6 + \chi _{2,8} z^8 + \chi _{2,10}z ^{10} ) \delta ^2  \\
	& \quad + ( \chi _{4,0} + \chi _{4,2} z ^2 + \chi _{4,4} z ^4 + \chi _{4,6} z^6 + \chi _{4,8}  z ^8 ) \delta ^4 + ( \chi _{6,0} + \chi _{6,2} z ^2 + \chi _{6,4} x ^4 + \chi _{6,6}z ^6 ) \delta ^6 \\
	& \quad + ( \chi _{8,0} + \chi _{8,2} z ^2 + \chi _{8,4} z ^4 ) \delta ^8  + (\chi _{10,0} + \chi _{10,2}  z ^2 ) \delta ^{10} + \delta  ^{12} \\
	& \quad  + 2 \lambda z ( \phi _{0,0} + \phi_{0,2} \delta ^2 + \phi _{0,4} \delta ^4 + \phi _{0,6} \delta ^6 )+ ( \phi _{2,0} + \phi _{2,2} \delta ^2  + \phi _{2,4} \delta ^4 ) z ^2 \\
	& \quad  + ( \phi _{4,0}+  \phi _{4,2} \delta ^2 ) z ^4 + z ^6 ) +2 \mu \delta ( \psi _{0,0}  + \psi _{0,2} z ^2 + \psi _{0,4} z ^4  + \psi _{0,6} z ^6 )  \\
	& \quad + ( \psi _{2,0}  + \psi _{2,2 }z ^2 + \psi _{2,4 } z ^4 ) \delta ^2 + ( \psi _{4,0} + \psi_{4,2} z ^2 ) \delta ^4 + \delta ^6 ) + ( \lambda ^2 + \mu ^2) \mathcal{R}_{1} (\delta,z), \label{eq124}
	\end{split}
	\end{align}
	where the form of $\mathcal{R}_{1}$ is as given in Eq. (\ref{eq4}). On substituting the above equation (\ref{eq124}) into the bilinear form (\ref{eq32}), we get a polynomial in $z$ and $\delta$. Equating all terms arising as the coefficient of the different powers of $z$ and $\delta$ separately, we obtain a system of nonlinear equations. On solving those equations, we arrive at the following third-order rogue wave parameters:
	\bes\bea 
	&& \chi _{0,0} = {\alpha  K  \left(878826025 \alpha ^5 \beta ^2 K^5 L^2+ 27 (\beta ^7 L^7 - \gamma ^7) \lambda ^2 \right)}\big/{9 \beta ^8 L^8 }, \\
	&& \chi _{0,2} = -\left({300896750  \alpha ^5 \beta ^2  K ^5 L ^2 + 3 (\beta ^7 L^7 - \gamma ^7) \lambda ^2}\right)\big/{3 \beta ^6 \gamma L^6  }, \\
	&& \chi _{0,4} = {16391725 \alpha ^4  K ^4 }\big/{3 \beta ^2 \gamma ^2 L^2 }, \quad
	\chi _{0,6}  = {-798980 \alpha ^3  K ^3 }\big/{3 \gamma ^3 }, \\
	&& \chi _{0,8} ={4335 \alpha ^2 \beta ^2  K ^2 L^2 }\big/{\gamma ^4 },\quad
	\chi _{0,10} = {-58 \alpha \beta^4 K  L^4} \big/{\gamma^5 }, \\
	&& \chi _{0,12} = {\beta ^6 L^6 }\big/{\gamma ^6 }, \quad
	\chi _{2,0}  = {-\left(3 (\beta ^7 L^7 - \gamma ^7) \lambda ^2+159786550 \alpha^5 \beta ^2  K ^5 L^2\right)}\big/{3 \beta ^7 L ^7 },\\
	&& ~\chi _{2,2} = { 565950 \alpha ^4  K  ^4  }\big/{\beta ^3 \gamma L^3},\quad
	\chi _{2,4}  = {14700 \alpha ^3  K ^3 }\big/{\beta \gamma ^2 L},\\
	&& \chi _{2,6 } = { 35420 \alpha ^2 \beta  K ^2 L }\big/{\gamma ^3 }, \quad
	\chi _{2,8}  = {-570 \alpha \beta ^3  K  L^3 }\big/{\gamma ^4 },\\
	&& \chi _{2,10} = {6 \beta ^5 L ^5 }\big/{\gamma ^5 }, \quad
	\chi_{4,0}  = {-5187875 \alpha ^4  K ^4 }\big/{3 \beta ^4 L^4}, \\
	&& \chi _{4,2}= {- 220500 \alpha ^3  K ^3  }\big/{\beta ^2 \gamma L^2}, \quad 
	\chi _{4,4} = {37450 \alpha ^2  K ^2 }\big/{\gamma ^2 }, \\
	&& \chi _{4,6} = {-1460 \alpha \beta ^2  K  L^2 }\big/{\gamma ^3 },~ 
	\chi_{4,8} = {15 \beta ^4 L^4}\big/{\gamma ^4 },~
	\chi _{6,0} = {-75460 \alpha ^3  K ^3 }\big/{3 \beta ^3 L^3}, \\
	&& \chi _{6,2} = {18620\alpha ^2  K ^2 }\big/{\beta \gamma L},\quad
	\chi _{6,4} = {-1540 \alpha \beta  K L }\big/{\gamma ^2},\quad
	\chi_{6,6}  = {20 \beta ^3 L^3}\big/{\gamma ^3 },\\
	&& \chi _{8,0} = {735 \alpha ^2  K  ^2 }\big/{\beta ^2 L^2},\quad
	\chi _{8,2}  = {-690 \alpha  K }\big/{\gamma },\quad
	\chi _{8,4} = {15 \beta ^2 L^2 }\big/{\gamma ^2 }, \\
	&& \chi _{10,0} = {-98 \alpha  K }\big/{\beta L},\quad
	\chi _{10,2}  = {6 \beta L}\big/{\gamma},\quad
	\phi _{0,0} = {-18865 \alpha ^3  \gamma ^3  K ^3 }\big/{3 \beta ^6 L^6 }, \\
	&& \phi _{0,2}= {-665 \alpha ^2 \gamma ^3 K^2 }\big/{\beta ^5 L^5 }, \quad
	\phi _{0,4}  = {-105 \alpha \gamma ^3  K  }\big/{\beta ^4 L^4 }, \quad
	\phi _{0,6} = {5 \gamma ^3}\big/{\beta ^3 L^3},\\
	&& \phi _{2,0} = {-245 \alpha ^2 \gamma ^2  K ^2 }\big/{\beta ^4 L^4}, \quad
	\phi _{2,2}  = {190 \alpha \gamma ^2  K }\big/{\beta ^3 L^3 },\quad
	\phi _{2,4}  = {-5 \gamma ^2 }\big/{\beta ^2 L^2 }, \\
	&& \phi _{4,0 } = {7 \alpha \gamma  K }\big/{\beta ^2 L^2},\quad
	\phi _{4,2} = {-9 \gamma }\big/{\beta L},\quad
	\psi_{0,0} = {-12005 \alpha ^3  K ^3}\big/{3 \beta^3 L^3}, \\
	&& \psi _{0,2} = {535 \alpha ^2  K ^2  }\big/{\beta \gamma L},\quad
	\psi _{0,4}  = {-45 \alpha \beta  K L}\big/{\gamma ^2 },\quad
	\psi _{0,6} ={5 \beta ^3 L^3}\big/{\gamma ^3}, \\
	&& \psi _{2,0} = {-245 \alpha ^2  K ^2 }\big/{\beta ^2 L^2},\quad
	\psi _{2,2} = {230\alpha K }\big/{\gamma},\quad
	\psi _{2,4} = {-5 \beta ^2 L^2 }\big/{\gamma ^2 }, \\
	&& \psi _{4,0} = {-13 \alpha K  }/{\beta L}, \quad
	\psi _{4,2} = {-9 \beta L}/{\gamma}.
	\eea \label{3opar}\ees
	Finally, using the above rogue wave parameters (\ref{3opar}), the form of $\mathcal{R}_3$ (\ref{eq124}), and the bilinearizing transformation $u(\delta,z)=u_0+2  (ln [\mathcal{R}_{3}] )_\delta$, the third-order rogue wave solution of the generalized (3+1)-dimensional Hirota-Satsuma-Ito equation (\ref{eq2a}) is obtained in the following form: 
	\begin{equation}
	u(x,y,z,t)=u_0+\dfrac{G_3(x,y,z,t)}{F_3(x,y,z,t)}, \label{eq13a}
	\end{equation}
	where $G_3(x,y,z,t)$ and $F_3(x,y,z,t)$ are polynomials of degree 11 and 12, respectively, and their exact expressions are given in appendix \ref{g3f3eqs}. 
	\begin{figure}[h]
		\centering\includegraphics[width=0.9\columnwidth]{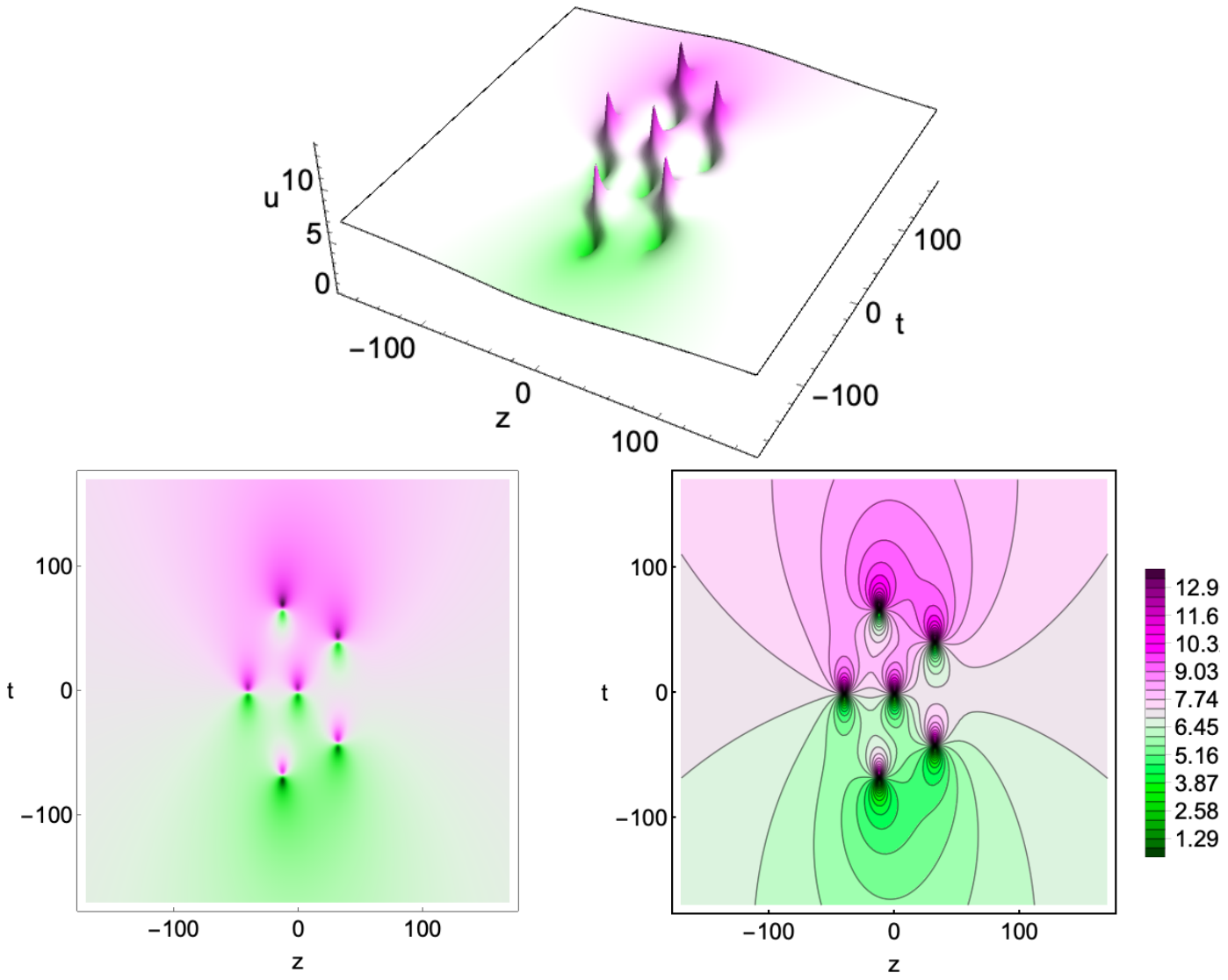}
		\caption{Nature of the third-order rogue wave forming a pentagon pattern with six separate (sextuple) doubly-localized up-down peak structures for the choice $u_0=7.5$, $\alpha=0.02$, $\beta=-0.09$, $\gamma=-3.9$, $\delta_1=0.2$, $\delta_2=0.1$, $\lambda=0.1$ and $\mu=-0.6$ with $y=0.1$ and $x=0.1$. The bottom panels show the two-dimensional density and contour plots.}
		\label{fig-third}
	\end{figure} 
	
\begin{figure}[h]
		\centering\includegraphics[width=0.92465\columnwidth]{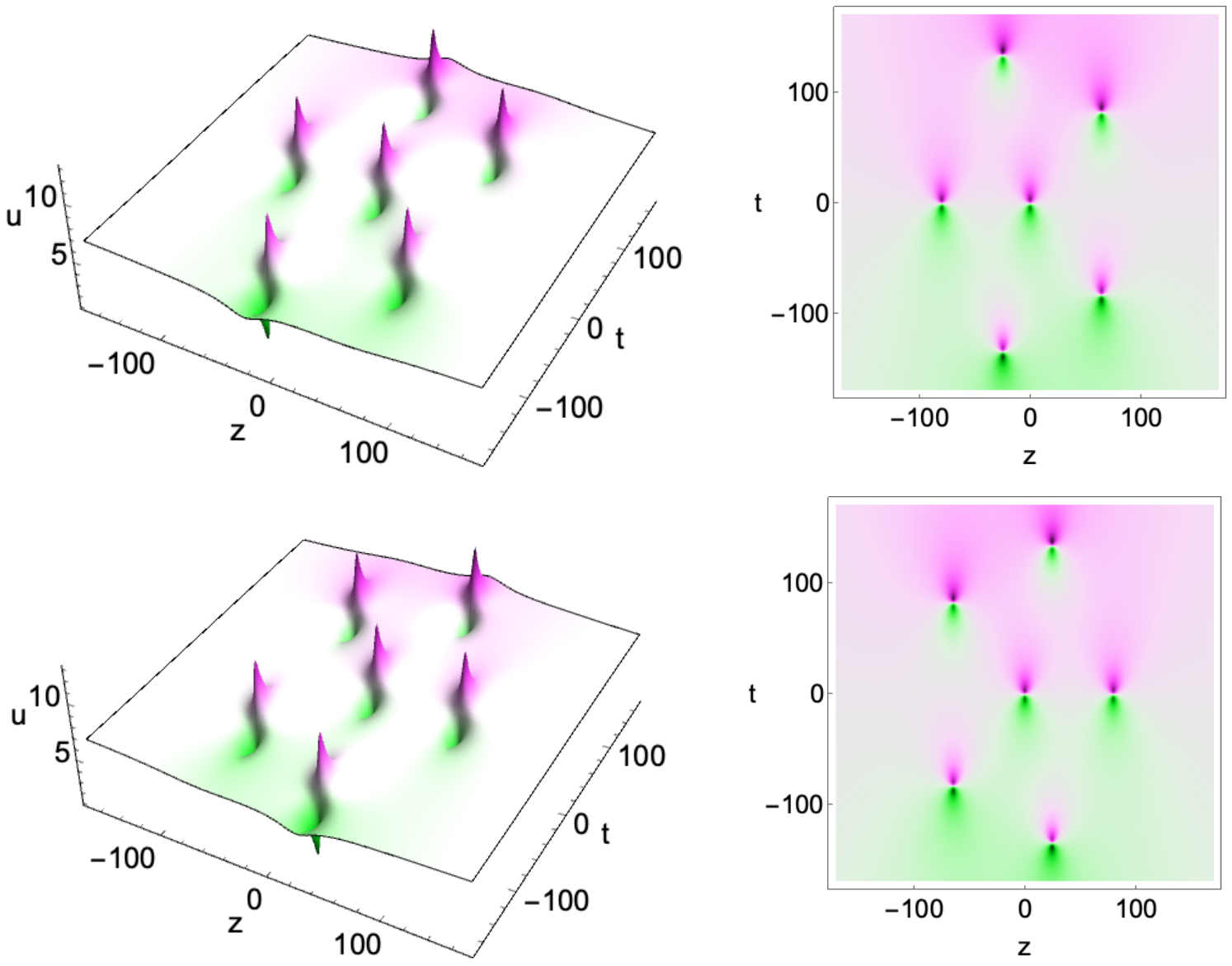}
		\caption{Rotation and spreading of the third-order rogue waves having sextuple (six-peak/dip) structure for two different choices of $\lambda$. Spreading of sextuple structure for $\lambda=3.1$ (top) and their rotated form with reflection symmetry along $z$ when $\lambda=-3.1$ (bottom) with other parameters same as given in Fig. \ref{fig-third}.}
		\label{fig-third2}
	\end{figure}

\begin{figure}[h]
		\centering\includegraphics[width=0.945\columnwidth]{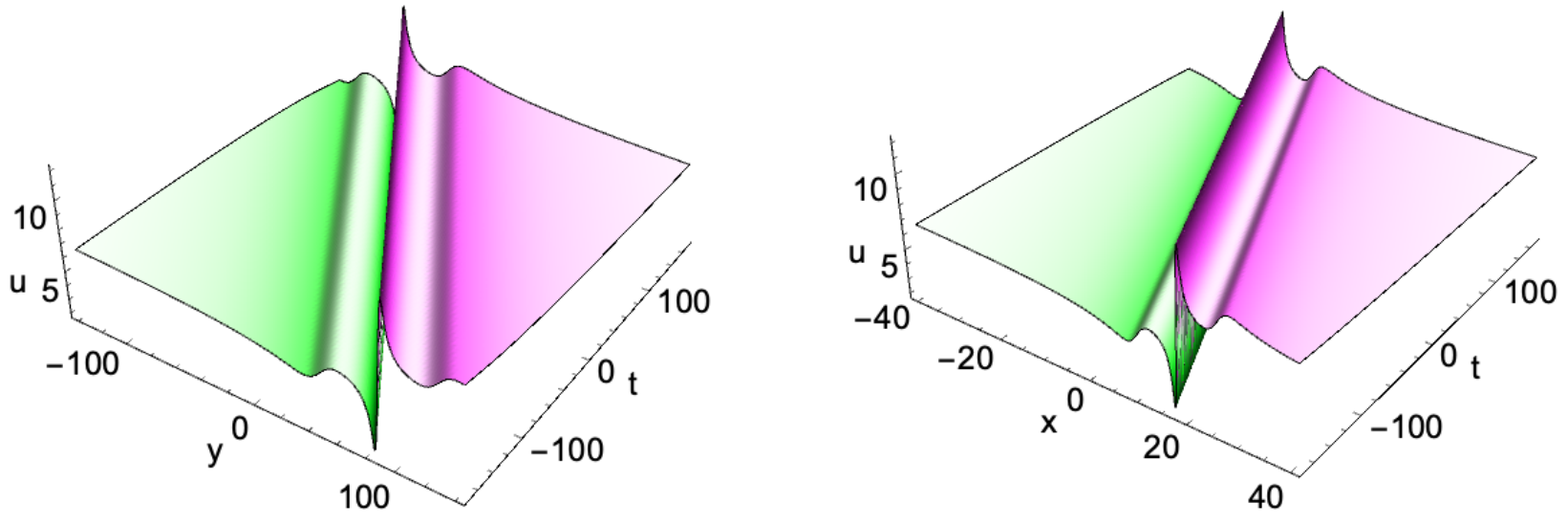}
		\caption{Third-order line rogue waves admitting three peaks and dips in $y-t$ for $z=0.1$ and $x=0.1$ (a) and in $x-t$ for $z=0.1$ and $y=0.1$ by keeping the other  parameters similar to Fig. \ref{fig-third}.}
		\label{fig-third3}
	\end{figure}

	Checking at the above third-order rogue wave solution (\ref{eq13a}), one can identify there are eight arbitrary parameters available to control its dynamics. Actually, it is lesser by two parameters than that of second-order because here all $\chi_{i,j}$, $\phi_{i,j}$ and $\psi_{i,j}$ are not arbitrary and become function of other parameters $\alpha$, $\beta$, $\gamma$, $\delta_1$, and $\delta_2$. Additionally, there exists $u_0$ which determines the required background amplitude and does not alters the nature or dynamics of the resultant rogue waves. It is important to note that the present third-order solution (\ref{eq13a}) reveals an interesting six-peak (sextuple) profile forming a pentagon with a center in the $z-t$ plane as shown in Fig. \ref{fig-third} where the identities of each doubly-localized structure seems symmetric. As discussed in the first- and second-order rogue waves, here also the identities of six-peak rogue wave can be manipulated based on the requirement with suitably chosen choices of available (eight) arbitrary parameter and their roles are much similar to those given in Sec. 3.2 and 3.3. Mainly, the amplitude, length, width, orientation, inter-peak/dip distance are able to be controlled by them. For completeness, we have shown such sextuple rogue waves and their rotation behaviour along the plane $z-t$ just by tuning $\lambda$ parameter in Fig. \ref{fig-third2}. Apart from these sextuple doubly-localized structures, the third-order rogue wave solution (\ref{eq13a}) also possesses singly-localized line rogue waves with triple-peaks and triple-dips along the planes $y-t$ and $x-t$ as demonstrated in Fig. \ref{fig-third3}. Similar to that of in the $z-t$ sextuple structure these line rogue waves can be altered by tuning the arbitrary parameters.
	
	
	\subsection{General Rogue wave of order $N$}
	Proceeding further, one can construct higher-order (arbitrary order $N=r+1$) rogue wave solutions in a straightforward manner by adopting $r\geq 3$ in the generalized polynomial function and solving the resultant system of equations arising from the bilinear form systematically. Due to more complex mathematical forms and length of the manuscript, we have not given them in the present work. However, its expression can be written in a generalized form as given below.
	\begin{equation}
	u(x,y,z,t)=u_0+\dfrac{G_r(x,y,z,t)}{F_r(x,y,z,t)}, \label{eq14a}
	\end{equation}
	where $G_r(x,y,z,t)$ and $F_r(x,y,z,t)$ are multi-variable polynomials respectively.
	The above $N$-th order rogue wave also vanishes to constant background $u_0$ as  $\lim_{|x|,|y|,|z| \to \infty}u(x,y,z,t)=u_0$, which clearly shows that the solution decays to the desired arbitrary background $u_0$ along all the spatial directions. Because of the fact that the denominator polynomial is always one degree higher than that of numerator, which in turn confirms the localization of the rogue waves solutions. From a thorough analysis one can also witness the formation of interesting multi-peak/dip doubly-localized structures with much freedom to control their identities with suitable parameters. For example, the fourth-, fifth-, sixth-, etc. order rogue wave solution resulting for the choice $r=3$, $r=4$, $r=5$, etc. respectively can consist of 10, 15, 21, etc. number of doubly-localized peaks \cite{yang21}, where their manipulation will be of much significance with the availability of extensive arbitrary parameters.

\section{Discussion} 	
 {For a better understanding of the manuscript, in this section, we highlight the motivation, significance of the study, important observations/results, novelty of the problem ad results compared to previous works along with certain possible future directions for immediate attention.}

\begin{itemize}
    \item  {The considered (3+1)-dimensional Hirota-Satsuma-Ito equation describing the dynamics of shallow water (\ref{eq13}) is new and more generalized one with different classes of nonlinear equations can be reduced for suitable $\Gamma_j$ parameters. Especially, it consists of nine different nonlinear soliton equations, including (2+1)D HSI equations, generalized Calogero-Bogoyavlenskii-Schiff equation, dimensionally reduced Jimbo-Miwa equation, (3+1)D generalized KP/BKP equations and other significant nonlinear wave equations.}

    \item  {{The integrability nature of the considered (3+1)D HSI model (\ref{eq13}) is studied by testing the Painlev\'e analysis. Nevertheless, it is found that the general model is non-integrable. However, interestingly, we have identified that a sub-case of the model with three arbitrary parameters for the choice $\Gamma_1=\Gamma_3=\Gamma_6=a_1,  \Gamma_2=\Gamma_5=\Gamma_7=a_2, \Gamma_8=0$, $\Gamma_4=a_3$ is arriving as Painlev\'e integrable, which is a new integrable soliton model. It is worth pointing out that we can still search for several model parameters, where (\ref{peq6}) vanish, and perform the Painlev\'e analysis to those cases, to find the integrable water waves models. }}

    \item  {{After observing the non-integrable nature of the (3+1)D HSI model (\ref{eq13}), we have obtained the higher-order rogue wave solutions and studied their evolutionary dynamics. It is an established fact that the rogue waves are considerably well localized nonlinear structures and their emergence is observed in different physical systems like deep ocean, shallow water, plasma, Bose-Einstein condensate, and optical models. Rogue waves are very volatile/chaotic nonlinear waves structures, which can causes severe damages in the associated systems and several reports are available in the literature. Construction of rogue waves are quite obvious for only integrable models, while the efforts for rogue waves to non-integrable models are very challenging task and impossible in several occasions. Because, the non-availability of Lax pair reduces the possibility of utilization of the techniques like Darboux transform, Gauge transform and Inverse spectral transform, to non-integrable equations and especially to higher-dimensional soliton models.}} 

    \item  {{ In such situation, the methodology adopted in this work offers a concrete and unified mechanism to obtain a generalized higher-order rogue wave solutions even to the non-integrable models. This methodology includes the Hirota bilinear formalism and generalised polynomial functions for the Hirota functions. Proceeding from every step of series parameters $r$ give order we can obtain the required $r+1$-th order wave solution which enable us to study the dynamics of such localized patterns in detail. The direct applicability of the Hirota bilinear method to obtain explicit rogue waves is comparatively less studied and requires more attention. If the model admits $N$-soliton solutions (which is another route for integrability) then its reduction can provide rogue waves in an alternative way and it will be considered as a future assignment.}}

    \item  {{The constructed rogue wave solutions given in the previous section for the (3+1)D HSI model (\ref{eq13}) shows that their evolution can be controlled/engineered with the help of arbitrary parameters. The first-order rogue wave shows doubly-localized up-down single-peak in $z-t$ plane and line rogue wave patterns in $x-t$ and $y-t$ planes with different choices over the arbitrary parameters. The properties of rogue waves such as amplitude, width, up-down peak spacing, etc. can be controlled appropriately tuning the $\delta_1$ and $\delta_2$ parameters as demonstrated in the previous section.}} 
    
    \item  {{In contrast to the first-order case, the second-order rogue wave describes the three separate doubly-localized up-down peaks, a design similar to the rogue wave triplet and forms a triangular shape. Another interesting fact is that we are able to rotate the rogue waves by using the $\lambda$, $\mu$, $\phi_{2,0}$, and $\psi_{2,0}$ parameters as depicted in Figs.  \ref{fig-second2} and \ref{fig-second3}. Further, we can also form a degenerate type (merged) first-order-like rogue wave structure as shown in Fig. \ref{fig-second4}. Also, its evolution exhibits second-order line rogue waves. }} 
    
    \item  {{Meanwhile, for third-order rogue wave solution, we have sextuple (six-peak/dip) structure forming a pentagon along with the third-order line rogue waves. We observed multi rogue waves with controllable patterns, having a sufficient number of arbitrary parameters to control its mechanism with eight arbitrary parameters in third, ten arbitrary parameters in second and six arbitrary parameters in the first-order rogue wave solution, which helps immensely to control amplitude, length, width, inter-peak/dip distances, and orientations. One can extend the analysis further in a straightforward manner to obtain any higher-order rogue wave solution as explained above and their dynamics can be explored.}}
    
    \item  {{Another future direction from the present study is to look for other nonlinear wave solutions to the present general non-integrable (3+1)D HSI model (\ref{eq13}) such as solitons, breathers, interaction waves and their coexisting dynamics can be studied. Additionally, the newly identified Painlev\'e integrable equation (\ref{eq130}) can be considered in a broader perspective to check its Lax pair using which further solutions and dynamics of various nonlinear waves can be investigated. }}

\end{itemize}
	
	\section{Conclusion} 
	In this work, we have considered an extended version of the Hirota-Satsuma-Ito equation describing the dynamics of shallow water waves in (3+1)-dimensions, which can be reduced to several known models including (2+1)-dimensional Hirota-Satsuma-Ito equation, Calogero-Bogoyavlenskii-Schiff, KP, BKP and Jimbo-Miwa equations, and shown that it does not pass the Painlev\'e test for integrability. Next, we have constructed an exact analytical form of rogue wave solutions. Particularly, we have established the working methodology to derive higher-order rogue wave solutions of arbitrary order ($N=r+1,~ r=0,1,2,3, \dots$) through the Hirota bilinear formalism and a generalized polynomial series. From the derived explicit analytical rogue wave solutions, we have carried out a detailed analysis and identified different pattern formation mechanism resulting from the advantages of several arbitrary parameters, that enable one to engineer the rogue waves based on the required properties. These patterns include the much impactful (spatio-temporal) doubly-localized rogue waves with multiple peak-dip structures and different orientations along the $z-t$ plane. Especially, we have portrayed that the first-, second-, and third-order rogue waves possess single, triple (triangular), and sextuple peak(s)/dip(s) peaks, respectively along $z-t$ plane with possibilities to control their identities such as amplitude, length, width, inter-peak/dip distances, and orientations by suitably tuning the arbitrary parameters. Further, the second-, and third-order rogue waves form triangular and pentagon type  geometrical patterns along $z-t$ plane. However, the obtained solutions exhibit spatially/singly-localized line-rogue waves along $x-t$ and $y-t$ planes, where they admit one-, two-, and three-peak(s)-dip(s) travelling line rogue waves that can also be controlled by changing the parameters. Every outcome of the analyses are graphically demonstrated appropriately for a clear understanding and completeness. The present work will be an important contribution to rogue wave dynamics in higher-dimensional nonlinear systems, including non-integrable nonlinear models.\\ 
	
	\setstretch{1.0}
	\noindent{\bf Acknowledgements}\\ 
	One of the authors Sudhir Singh would like to thank the National Institute of Technology Tiruchirappalli and the Ministry of Human Resource Development, Govt. of India, for the financial support through institute fellowship. The research work of K. Sakkaravarthi was supported by the Korean Ministry of Education Science and Technology through Young Scientist Training (YST) Program of the Asia-Pacific Center for Theoretical Physics (APCTP), Pohang-si, Gyeongsangbuk-do. K. Sakkaravarthi was also partially supported by Department of Science and Technology - Science and Engineering Research Board (DST-SERB), Govt. of India, sponsored National Post-Doctoral Fellowship (File No. PDF/2016/000547). The authors also thank the anonymous reviewers for the fruitful comments and suggestions.\\
	
	\noindent{\bf Declaration}\\
	The authors declare that there is no conflict of interests regarding the research effort and the publication of this manuscript.\\
	
	\noindent{\bf CRediT Author Contribution Statement}\\ {\bf Sudhir Singh}: Conceptualization, Methodology, Writing - Original Draft Preparation, Writing - Review \& Editing. {\bf K. Sakkaravarthi}: Validation, Formal Analysis, Investigation, Visualization, Writing - Original Draft Preparation, Writing - Review \& Editing. {\bf T. Tamizhmani}: Resources, Writing - Review \& Editing. {\bf K. Murugesan}: Resources, Writing - Review \& Editing, Funding acquisition, Supervision. 
	
	\appendix
	
	\section{Explicit expression for third-order rogue wave solution (\ref{eq13a})} \label{g3f3eqs}
	The explicit form of $G_3(x,y,z,t)$ and $F_3(x,y,z,t)$ obtained for the third-order rogue wave solution (\ref{eq13a}) are as follows.
	\bes\bea 
	&G_3=&
	12{L}\beta \gamma  (18{L}^{12}{z}^{10}\beta^{12}({x}+ \delta _1 y+\delta _2 t)+90{L}^{{l}1}{z}^{8}\beta^{11}\gamma ({x}+ \delta _1 y+\delta _2 t)^{3}\nonumber\\ &&+3\gamma ^{12}({x}+ \delta _1 y+\delta _2 t)\lambda^{2}-70{K}{L}^{3}\alpha\beta^{3}\gamma ^{5}({x}+ \delta _1 y+\delta _2 t)^{3}(148225{K}^{3}\alpha^{3}+18{z}\gamma ^{3}\lambda) \nonumber\\ &&- 70{K}^{2}{L}^{2}\alpha^{2}\beta^{2} \gamma^{5}({x}+ \delta _1 y+\delta _2 t)(2282665{K}^{3}\alpha^{3}+57{z}\gamma ^{3}\lambda) \nonumber\\ &&+ 15{L}^{5}\beta^{5}\gamma^{4}({x}+ \delta _1 y+\delta _2 t)^{2}(-88{20} 0{K}^{3}{z}^{2}\alpha^{3}({x}+ \delta _1 y+\delta _2 t)- 4 {z}^{3}\gamma^{3}({x}+ \delta _1 y+\delta _2 t)\lambda \nonumber\\ &&+147{K}^{2}\alpha^{2}\gamma (4({x}+ \delta _1 y+\delta _2 t)^{5}-\mu))+ 15 {L}^{9}{z}^{4}\beta^{9}\gamma ^{2}({x}+ \delta _1 y+\delta _2 t)^{2}(-584{K}{z}^{2}\alpha({x}+ \delta _1 y+\delta _2 t)\nonumber\\ &&+12 {\gamma }({x}+ \delta _1 y+\delta _2 t)^{5}-3{\gamma }\mu)+ 3{L}^{7}\beta^{7}\gamma ^{3}({x}+ \delta _1 y+\delta _2 t)(74900{K}^{2}{z}^{4}\alpha^{2}({x}+ \delta _1 y+\delta _2 t)^{2} \nonumber\\ &&-690{K}{z}^{2}\alpha \gamma ({x}+\delta _1 y+\delta _2 t) (4({x}+\gamma\delta _1+{t}\delta _2)^{5}-\mu)+\gamma ^{2}(({x}+ \delta _1 y+\delta _2 t)^{5}+\mu)(6({x}+ \delta _1 y+\delta _2 t)^{5}+\mu)) \nonumber\\ &&- 15 {L}^{10}{z}^{6}\beta^{10}\gamma (114{K}{z}^{2}\alpha({x}+ \delta _1 y+\delta _2 t)-\gamma (12({x}+ \delta _1 y+\delta _2 t)^{5}+\mu))\nonumber\\ 
	&&+ 15 {L}^{8}{z}^{2}\beta^{8}\gamma^{2} (7084\ {K}^{2}{z}^{4}\alpha^{2}({x}+ \delta _1 y+\delta _2 t)+3 {\gamma ^{2}}({x}+ \delta _1 y+\delta _2 t)^{4} (2({x}+ \delta _1 y+\delta _2 t)^{5}-3\mu)\nonumber\\ &&-3{K}{z}^{2}\alpha \gamma (308({x}+ \delta _1 y+\delta _2 t)^{5}+3\mu))+ 5 {L}^{4}\beta^{4}\gamma ^{4}(339570{K}^{4}{z}^{2}\alpha^{4}({x}+ \delta _1 y+\delta _2 t)\nonumber\\ &&+228{K}{z}^{3}\alpha \gamma ^{3}({x}+ \delta _1 y+\delta _2 t)\lambda+ 18\ {z}\gamma ^{4}({x}+ \delta _1 y+\delta _2 t)^{5}\lambda-343{K}^{3}\alpha^{3}\gamma (132({x}+ \delta _1 y+\delta _2 t)^{5}+7\mu)) \nonumber\\ &&-  3 {L}^{6}\beta^{6}\gamma ^{3}(-14700{K}^{3}{z}^{4}\alpha^{3}({x}+ \delta _1 y+\delta _2 t)+{K}\alpha \gamma ^{2}({x}+ \delta _1 y+\delta _2 t)^{4} (490({x}+ \delta _1 y+\delta _2 t)^{5}+13\mu)\nonumber\\ &&-5{K}^{2}{z}^{2}\alpha^{2}\gamma \ (11172\ ({x}+ \delta _1 y+\delta _2 t)^{5}+107\mu)+ 2 \gamma ^{2}({x}+ \delta _1 y+\delta _2 t)(9{z}^{5}\gamma {\lambda}+26\alpha {K}({x}+ \delta _1 y+\delta _2 t)^{3}\mu))),\nonumber
	\eea
	\bea 
	&F_3=&
	(9{L}^{14}{z}^{12}\beta^{14}+54{L}^{13}{z}^{10}\beta^{13}\gamma({x}+ \delta _1 y+\delta _2 t)^{2}+ 9 {L}^{12}{z}^{8}\beta^{12} \gamma \nonumber\\ &&(-58{K}{z}^{2}\alpha+15\gamma ({x}+ \delta _1 y+\delta _2 t)^{4})-27{K}\alpha \gamma ^{13}\lambda^{2}+ 9 {L}\beta \gamma ^{13}({x}+ \delta _1 y+\delta _2 t)^{2}\lambda^{2}\nonumber\\ &&-210{K}^{2}{L}^{3}\alpha^{2}\beta^{3}\gamma ^{6}({x}+ \delta _1 y+\delta _2 t)^{2}(2282665{K}^{3}\alpha^{3}+57{z}\gamma ^{3}\lambda)\nonumber\\ &&- 105{K}{L}^{4}\alpha\beta^{4}\gamma ^{5}(8597050{K}^{4}{z}^{2}\alpha^{4}+148225{K}^{3}\alpha^{3}\gamma ({x}+ \delta _1 y+\delta _2 t)^{4}+42{K}{z}^{3}\alpha\gamma^{3}\lambda \nonumber\\ &&+ 18\ {z}\gamma ^{4}({x}+ \delta _1 y+\delta _2 t)^{4}\lambda)+{L}^{2}\beta^{2}\gamma ^{6}(878826025{K}^{6}\alpha^{6}-113190{K}^{3}{z}\alpha^{3}\gamma ^{3}\lambda+9{z}^{2}\gamma ^{6}\lambda^{2}) \nonumber\\ &&+ 3{L}^{6}\beta^{6}\gamma ^{4}(16 391725 {K}^{4}{z}^{4}\alpha^{4}-661500{K}^{3}{z}^{2}\alpha^{3}\gamma ({x}+ \delta _1 y+\delta _2 t)^{4}+42{K}{z}^{5}\alpha \gamma ^{3}\lambda \nonumber\\ &&- 30\ {z}^{3}\gamma ^{4}({x}+ \delta _1 y+\delta _2 t)^{4}\lambda+735{K}^{2}\alpha^{2}\gamma ^{2}({x}+ \delta _1 y+\delta _2 t)^{3}(3({x}+ \delta _1 y+\delta _2 t)^{5}-2\mu))\nonumber\\ &&+ 90 {L}^{11}{z}^{6}\beta^{11}\gamma ^{2}({x}+ \delta _1 y+\delta _2 t) (-57{K}{z}^{2}\alpha({x}+ \delta _1 y+\delta _2 t)+\gamma (2({x}+ \delta _1 y+\delta _2 t)^{5}+\mu)) \nonumber\\ &&+ 30 {L}^{5}\beta^{5}\gamma ^{5}(x+y \delta _1+{t}\delta _2) (169785{K}^{4}{z}^{2}\alpha^{4}({x}+ \delta _1 y+\delta _2 t)+114{K}{z}^{3}\alpha \gamma ^{3}({x}+ \delta _1 y+\delta _2 t) \lambda \nonumber\\ &&+3{z}\gamma ^{4}({x}+ \delta _1 y+\delta _2 t)^{5}\lambda-343{K}^{3}\alpha^{3}\gamma(22({x}+ \delta _1 y+\delta _2 t)^{5}+7\mu)) \nonumber\\ &&+ 45 {L}^{10}{z}^{4}\beta^{10}\gamma ^{2}(867{K}^{2}{z}^{4}\alpha^{2}-292{K}{z}^{2}\alpha \gamma ({x}+ \delta _1 y+\delta _2 t)^{4}+ \gamma ^{2}(3({x}+ \delta _1 y+\delta _2 t)^{8}
	\nonumber\\ &&-2({x}+ \delta _1 y+\delta _2 t)^{3}\mu))+9{L}^{9}{z}^{2}\beta^{9}\gamma ^{3} (35420 {K}^{2}{z}^{4}\alpha^{2}({x}+ \delta _1 y+\delta _2 t)^{2}\nonumber\\ &&-10{K}{z}^{2}\alpha \gamma ({x}+ \delta _1 y+\delta _2 t) (154({x}+ \delta _1 y+\delta _2 t)^{5}+9\mu)+ \gamma ^{2}(6{x}^{10}+60{x}^{9}( \delta _1 y+\delta _2 t)\nonumber\\ &&+270{x}^{8}(y \delta _1+{t}\delta _2)^{2}+{720} {x}^{7}(y \delta _1+{t}\delta _2)^{3}+ 1260\ {x}^{6}(y \delta _1+{t}\delta _2)^{4}+60{x}(y \delta _1+{t}\delta _2)^{9}\nonumber\\ &&+6(\delta _1 y+\delta _2 t)^{10}+ 90\ {x}^{2}(y \delta _1+{t}\delta _2)^{3}(3(\delta _1 y+\delta _2 t)^{5}-2\mu)+180{x}^{3}(y \delta _1+{t}\delta _2)^{2}  (4(y \delta _1+{t}\delta _2)^{5}-\mu)\nonumber\\ &&+90{x}^{4}(y \delta _1+{t}\delta _2)(14(y \delta _1+{t}\delta _2)^{5}-\mu)+ 18\ {x}^{5}(84(y \delta _1+{t}\delta _2)^{5}-\mu)-90{x}(y \delta _1+{t}\delta _2)^{4}\mu \nonumber\\ &&-18(y \delta _1+{t}\delta _2)^{5}\mu+\mu^{2}))- 9 {L}^{7}\beta^{7}\gamma ^{4}(-14700{K}^{3}{z}^{4}\alpha^{3}({x}+ \delta _1 y+\delta _2 t)^{2}+18{z}^{5}\gamma ^{3}({x}+ \delta _1 y+\delta _2 t)^{2}\lambda \nonumber\\ &&- 10\ {K}^{2}{z}^{2}\alpha^{2}\gamma ({x}+ \delta _1 y+\delta _2 t)\ (1862\ ({x}+ \delta _1 y+\delta _2 t)^{5}+107\mu)+ {K}\alpha \gamma ^{2}(98({x}+ \delta _1 y+\delta _2 t)^{10} \nonumber\\ &&+26({x}+ \delta _1 y+\delta _2 t)^{5}\mu+3\mu^{2}))+ 3 {L}^{8}\beta^{8}\gamma ^{3}(-798980{K}^{3}{z}^{6}\alpha^{3}+112350{K}^{2}{z}^{4}\alpha^{2}\gamma ({x}+ \delta _1 y+\delta _2 t)^{4} \nonumber\\ &&- 690\ {K}{z}^{2}\alpha \gamma ^{2}({x}+ \delta _1 y+\delta _2 t)^{3}(3({x}+ \delta _1 y+\delta _2 t)^{5}-2\mu)\nonumber\\ &&+ 3\ \gamma ^{3}(2{z}^{7}\lambda+({x}+ \delta _1 y+\delta _2 t)^{2}(({x}+ \delta _1 y+\delta _2 t)^{5}+\mu)^{2}))),\nonumber
	\eea
	where \bea K=(\delta _1+\delta _2), \mbox{ and  }  L=(1+\delta _1+\delta _2+\delta _1^{2}+\delta _1\delta _2).\nonumber
	\eea
	\ees
	\section*{References}

\end{document}